\documentclass[a4paper,11pt]{article}
\usepackage{jcappub} 
\usepackage{natbib}
\usepackage{graphicx,color}

\usepackage[T1]{fontenc} 
\usepackage{lineno, booktabs}

\newcommand{\lsim}{\mathrel{\mathop{\kern 0pt \rlap
  {\raise.2ex\hbox{$<$}}}
  \lower.9ex\hbox{\kern-.190em $\sim$}}}
\newcommand{\gsim}{\mathrel{\mathop{\kern 0pt \rlap
  {\raise.2ex\hbox{$>$}}}
  \lower.9ex\hbox{\kern-.190em $\sim$}}}

\title{Dipole anisotropy in cosmic electrons and positrons: inspection on local sources }

\author[a,b]{S. Manconi}
\author[c]{M. Di Mauro,}
\author[a,b]{F. Donato}

\affiliation[a]{Dipartimento di Fisica, Universit\`a di Torino, via P. Giuria 1, 10125 Torino, Italy}
\affiliation[b]{Istituto Nazionale di Fisica Nucleare, Sezione di Torino, Via P. Giuria 1, 10125 Torino, Italy}
\affiliation[c]{W. W. Hansen Experimental Physics Laboratory, Kavli Institute for Particle Astrophysics and Cosmology, Department of Physics and SLAC National Accelerator Laboratory, Stanford University, Stanford, CA 94305, USA}

\emailAdd{manconi@to.infn.it}
\emailAdd{mdimauro@slac.stanford.edu}
\emailAdd{donato@to.infn.it}

  \abstract{
The cosmic electrons and positrons have been measured with unprecedented statistics up to several hundreds GeV, thus  permitting 
to explore the role that close single sources can have in shaping the flux at different energies. 
In the present analysis, we consider electrons and positrons in cosmic rays to be produced by spallations of hadron fluxes with the interstellar medium, 
by a smooth Supernova Remnant (SNR) population, by all the ATNF catalog pulsars, and by few discrete, local SNRs.
We test several source models on the $e^++e^-$ and $e^+$ AMS-02 flux data.  For the configurations compatible with the data, we compute 
the dipole anisotropy in $e^++e^-$, $e^+$, $e^+/e^-$   from single sources. 
Our study includes a dedicated analysis to the Vela SNR. 
We show that {\it Fermi}-LAT present data on dipole anisotropy of $e^++e^-$ start to explore some of the models for the 
Vela SNR selected by AMS-02 flux data. 
We also investigate how the observed anisotropy could result from a combination of local sources. 
Our analysis shows that the search of anisotropy in the lepton fluxes up to TeV energies 
can be an interesting tool for the inspection of properties of close SNRs, complementary to the high precision flux data. 
}

\begin{document}
\maketitle
\flushbottom

\section{Introduction}
\label{sec:intro}
The data on cosmic-ray (CR) electrons and positrons ($e^{+}$ and $e^{-}$) have achieved a high statistical accuracy and an unprecedented energy coverage. Precision data on the positron fraction has been recently provided by the Pamela \citep{2009Natur.458..607A}, {\it Fermi}-LAT \citep{2012PhRvL.108a1103A} and 
AMS-02 \citep{PhysRevLett.110.141102,PhysRevLett.113.121101} Collaborations, as well as measurements of the electron, positron  \citep{2014PhRvL.113l1102A,Adriani:2013uda,2011PhRvL.106t1101A}
and electron plus positron \citep{2014PhRvL.113v1102A,Abdo:2009zk} fluxes. 
\\
These leptonic data can be interpreted with the emission of electrons and positrons from Supernova Remnants (SNRs), Pulsar Wind Nebulae (PWNe) and the so called secondary production given by the spallation reactions of primary CRs with the nuclei of the interstellar medium (ISM). The accuracy of $e^{+}$ and $e^{-}$ data has permitted an in-depth study of properties of these different sources (see \textit{e.g.} \cite{Boudaud:2014dta, Jin:2014ica, Yuan:2013eja, Lin:2014vja, Mertsch:2014poa, Gaggero:2013nfa, Mlyshev:2009twa, 2014JCAP...04..006D, 2016JCAP...05..031D}).
\\
CR electrons and positrons detected at high-energy ($E>10$ GeV) are the result of local emission mechanisms, since the time scale of energy losses for leptons is  smaller than the diffusion one (see e.g. \cite{1998ApJ...509..212S,Delahaye:2008ua, Evoli:2016xgn}).
The inspection of close sources for $e^+$ and $e^-$ is therefore crucial for the interpretation of the data \citep{2009APh....32..140G,2011APh....34..528D, Hooper:2008kg, Kobayashi:2003kp}.
\\
Recently, the lepton data have also been analyzed in terms of their arrival direction. Indeed, no significant anisotropy has been detected, and upper limits on the dipole term $\Delta$ have been set for the electron plus positron flux \citep{2010PhRvD..82i2003A}, the positron to electron ratio \citep{PhysRevLett.110.141102} and the positron flux \citep{Adriani:2015kfa}. As we will demonstrate in this paper, the dipole anisotropy in the leptons 
arriving at the Earth can be a remarkable observable to constrain the properties of astrophysical emitters of $e^+$ and $e^-$.
Dipole upper limits have already been  used to examine discrete source properties (see e.g. \cite{1971ApL.....9..169S,2011APh....34..528D, Kobayashi:2003kp}) or to discriminate between dark matter and pulsar scenarios \citep{Buesching:2008hr, Pato:2010im, 2010APh....34...59C, 2009APh....32..140G}. 
\\
We revisit in this paper the predictions for the anisotropy from PWNe and SNRs using the very recent  AMS-02 data for $e^{+}$ and $e^++e^-$ energy spectra, as drivers to predict the corresponding dipole term.
The results are then compared to all existing upper limits from AMS-02, {\it Fermi}-LAT and Pamela.
We explore different scenarios, and in particular  the hypothesis of a dominant SNR that is the major contributor to the $e^-$ spectrum, and the case of a PWN that dominates the $e^+$ spectrum. We consider the most updated catalogs for PWNe and SNRs, selecting the closest and most powerful sources according to radio observations.
We also investigate a more realist scenario where all the close PWNe and SNRs contribute to the anisotropy,
providing a realistic prediction for the composite anisotropy as a function of the energy.
\\
The paper is organized as follows: in Sec.~\ref{sec:generics} we describe our model of propagation of CR electrons and positrons and their emission mechanism from PWNe, SNRs and the secondary production; in Sec.~\ref{sec:thframe} we explain our model for the estimation of the anisotropy and we list the different scenarios that we scrutinize in the paper; in Sec.~\ref{sec:fluxes} we present the fit to AMS-02  $e^{+}$ and $e^{-}$ data, that we use to fix the emission properties from the different components;  in Sec.~\ref{sec:anisotropy} we report the results for the anisotropy in the different cases, and we 
conclude in Sect.~\ref{sec:concl}.

\bigskip

\section{Electrons and positrons in the Milky Way}
\label{sec:generics} 
Electrons and positrons can be produced by different mechanisms and sources in the Galaxy. In particular, 
primary electrons can be injected in the ISM by a SNR following 
a first order Fermi acceleration, while $e^\pm$ pairs can be accelerated by pulsars (PSR) and released 
by the PWN in the ISM. 
A source for secondary electrons and positrons is nourished by the fragmentation of nuclei CRs (mainly proton and helium) on the ISM (H and He). 
After being injected in the Galaxy, electrons and positrons loose energy at a high rate while being randomly diffused by the inhomogeneities of the Galactic magnetic field. 
The number density $\psi = \psi(E, \mathbf{x}, t)\equiv dn/dE$ per unit volume and energy 
obeys the generic transport equation 
(see \cite{2010A&A...524A..51D} and references therein):
\begin{equation}
 \frac{\partial \psi}{\partial t}  - \mathbf{\nabla} \cdot \left\lbrace K(E)  \mathbf{\nabla} \psi \right\rbrace + 
 \frac{\partial }{\partial E} \left\lbrace \frac{dE}{dt} \psi \right\rbrace = Q(E, \mathbf{x}, t)
 \label{eq:diff}
\end{equation}
where $K(E)$ is the energy dependent diffusion coefficient, $dE/dt\equiv b(E)$ accounts for the energy losses and $Q(E, \mathbf{x}, t)$ includes all the possible electron and positron sources.
We consider the factorization of the source term so that $Q(E, \mathbf{x}, t)= Q(E)\cdot \rho(\mathbf{x}) \cdot f(t)$.
As for the results of this paper, the solution to Eq.~\ref{eq:diff} is found within a semi-analytical model in which the Galaxy is shaped as cylinder with 
a thin disk with half-height $h=100$ pc and a thick magnetic halo whose vertical extension is $L$ \citep{Maurin:2001sj}. 
The radius of the cylinder, whenever not set to infinite, is fixed to $R_{\rm disc} = 20$ kpc. The distance of the Solar System  to the 
Galactic center is fixed to $r_{\odot} = 8.33$~kpc \citep{Gillessen:2008qv}.
We assume a spatial independent diffusion coefficient
\begin{equation}
\label{eq:diff_coeff}
 K(E)= \beta K_0 (\mathcal{R}/1 \text{GV})^\delta \simeq K_0 (E/1 \text{GeV})^{\delta}
\end{equation}
where $\beta=v/c$ is the Lorentz factor (for relativistic electrons, as in this analysis, $\beta=1$)
and $\mathcal{R}$ is the particle rigidity. 
We have included energy losses of electrons by Inverse Compton  scattering off the interstellar radiation field, 
and the synchrotron emission on the Galactic magnetic field. The black body approximation for the interstellar photon populations
at different wavelengths has been taken from \cite{2010A&A...524A..51D} (model M2 in their Table 2). The Galactic magnetic field intensity 
has been assumed $B=3.6\; \mu$G, as resulting from the sum (in quadrature) of the regular and turbulent components \citep{2007A&A...463..993S}. 
The free parameters for the propagation sector are $K_0, \delta$ and $ L$. 
As benchmark models, we 
consider the MED and MAX sets of propagation parameters \citep{Donato:2003xg}. Indeed, the recent fits to boron--to--carbon, B/C, and antiprotons AMS-02 data (see e.g. \cite{Genolini:2015cta, Kappl:2015bqa}) points to values of $K_0, \delta$ and $ L$ consistent with these two models. We have checked that using $K_0, \,\delta$ and $ L$ as derived in \cite{Genolini:2015cta, Kappl:2015bqa} the electron fluxes are included between the fluxes obtained for the MED and MAX propagation set-ups.
\\
We model the CR electron and positron sources according to the framework developed in \cite{2010A&A...524A..51D} and  \cite{2014JCAP...04..006D,2016JCAP...05..031D}, for which we address for any details. 
In the following, we will remind the main features, and the slight differences introduced here. 
We will be interested both in the global source distribution, describing effectively the Galactic population with average parameters, and in the properties of single, nearby sources. 

For the spatial distribution of sources $\rho(r, z)$, we follow the usual factorization between the radial source density $\rho(r)$ and 
in the exponential vertical profile:
\begin{equation}\label{eq:rho}
 \rho(r, z) = \rho_0 \rho(r) \exp \left( - \frac{|z|}{z_0} \right)
\end{equation}
where $r$ is the distance from the Galactic center along the Galactic plane, $z$ is the vertical coordinate, and $z_0=h=0.1$~kpc. 
The coefficient $\rho_0$ is fixed to normalize to unity the spatial distribution $\rho(r,z)$ within the diffusion halo of radius $R_{\rm disc}$ and 
half thickness $L$. 
For the SNRs, we make use of the radial distribution (G15 hereafter) recently obtained by \cite{2015MNRAS.454.1517G} 
analyzing the SNR sample contained in the Green catalog:
\begin{equation}\label{eq:rhor}
 \rho(r) \propto \left( \frac{r}{r_{\odot}} \right)^{\alpha_1} \exp \left( - \alpha_2 \frac{(r-r_{\odot})}{r_{\odot}}\right)
 \end{equation}
where $\alpha_1=1.09$, $\alpha_2=3.87$. 
 \begin{figure}[t]
 \centering\includegraphics[width=0.6\textwidth]{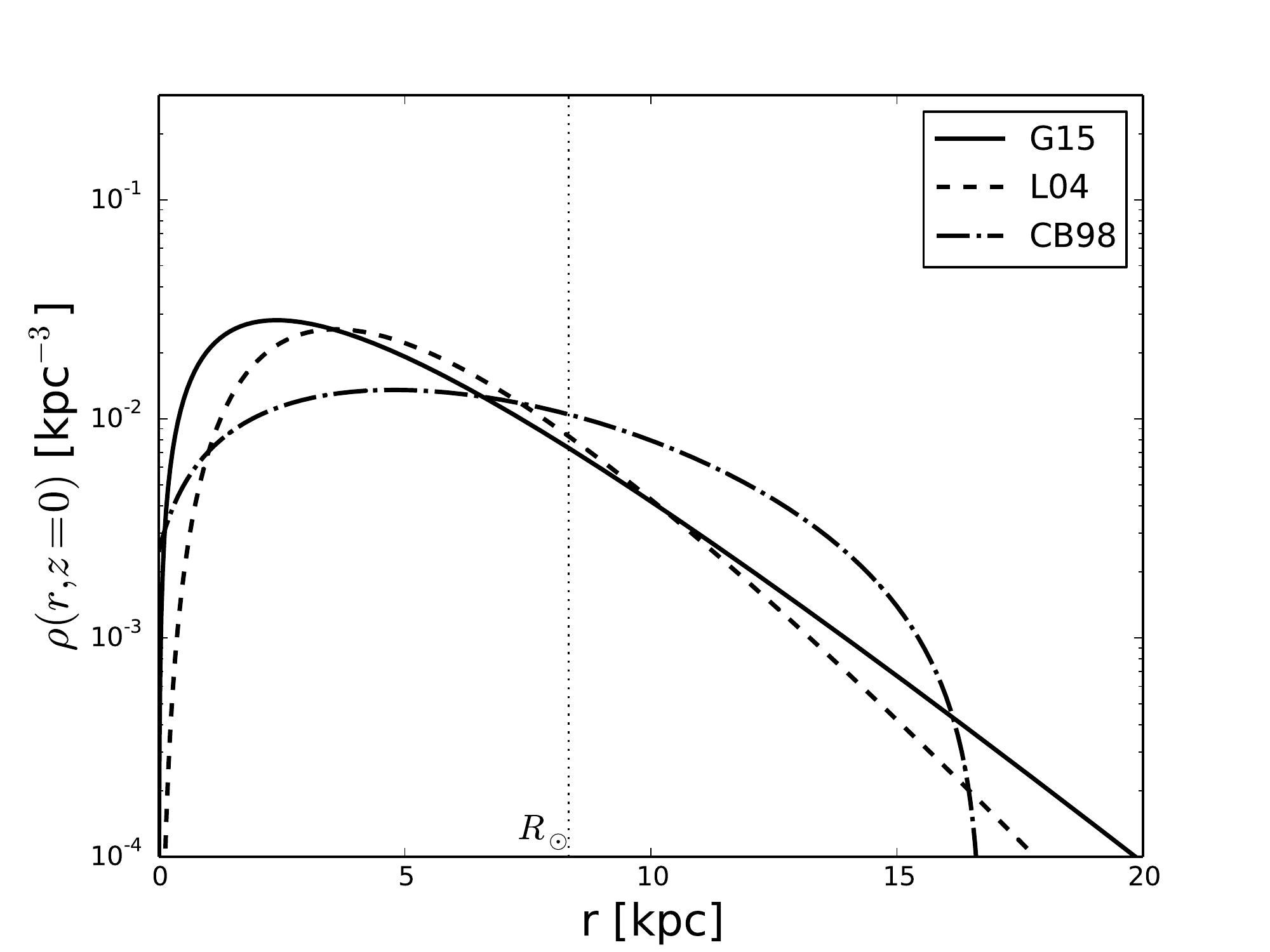}
 \caption{Radial distributions for SNRs and pulsars. The solid line corresponds to  
the G15  \cite{2015MNRAS.454.1517G} model, based on the brightest SNRs in the Green 
catalog; dashed line dubbed L04 is for the  \cite{2004IAUS..218..105L} distribution, built on pulsars 
from the ATNF catalog. Also shown is the model in \cite{1998ApJ...504..761C}. 
A black dotted line indicates the Earth position. }  \label{fig:rho}
\end{figure}
In Fig.~\ref{fig:rho}  we reproduce the radial profile G15, together with the largely used model 
obtained from pulsars in the ATNF catalog \citep{2004IAUS..218..105L} (L04 hereafter). 
For illustrative purposes, we also draw the model in \cite{1998ApJ...504..761C}, based on SNR 
observations and empirically scaled to compensate for observational selection effects. 
The L04 and G15 models are similar around the Solar System, while they show a remarkable difference 
toward the Galactic center. At variance,  the $\rho(r)$ proposed by \cite{1998ApJ...504..761C} is significantly flat for most of 
radii. Since electrons detected at the Earth have been produced by nearby (very few kpc) sources, we are mostly interested 
in the differences in the profiles around the Solar System. We will briefly discuss the effects of using  the radial distribution 
obtained with L04. 
Any time a smooth SNR population will be included in our analysis, it follows the radial profile in  Eq.~\ref{eq:rhor}. 
\\
We also consider the case of a smooth SNR population active only beyond a radius $R_{\rm cut}$ from the Earth (the {\it far} component), while the 
contribution inside $R_{\rm cut}$ (the {\it near} component) is given by the single sources as directly found in the Green 
SNR catalog. The cut in the smooth population is meant as a cylinder of radius 
 $R \equiv |r- r_{\odot}|\leq R_{\rm cut}$ and height $L$ around the Earth. 
This cylindrical Galactic portion is depleted of any smoothly distributed population. 
We have verified that our results are unmodified for a 3D spherical cut.  This is understood given the vertical distribution in Eq.~\ref{eq:rho}, where the exponential factor with $z_0=0.1$~kpc suppresses the source density outside the disk. 
\\
As for the PWNe, they are always  taken from the ATNF catalog. 
Their position in the Galaxy is set individually and picked from catalog. 
We  include only middle-aged pulsars, with observed age $50$~kyr$<t_{\rm obs}<10000$~kyr, since electron and positrons pairs accelerated to TeV energies in the termination shock (for a 
PWN review see  \cite{Gaensler:2006ua}) are believed to be confined  in the nebula or in the SNR until the merge of this system with the ISM, 
estimated to occur at least $40-50$~kyr after the pulsar formation \citep{1996ApJ...459L..83C, 2011ASSP...21..624B}. 
\\
As for the energy injection spectrum $Q(E)$, we adopt the function
 \begin{equation}
 Q(E)= Q_{0} \left( \frac{E}{E_0}\right)^{- \gamma} \exp \left(-\frac{E}{E_c} \right) 
 \label{eq:Q_E}
\end{equation}
for both SNR and PWN, where $Q_{0}$ is in units of GeV$^{-1}$ and $E_c$ is the cutoff energy. Throughout the text, we will label SNR or PWN the free parameters
which are expected to be different for the two distinct populations. If not differently stated, we adopt for both SNR and PWN $E_c= 5$~TeV.
The normalization of the power law is fixed to $E_0= 1$~GeV.
Given the injection spectrum in Eq.~\ref{eq:Q_E}, the total energy emitted in $e^-$ (for SNR) or $e^\pm$ (for PWN) in units of GeV (or erg) can be obtained as (see \cite{2010A&A...524A..51D})
\begin{equation}
 E_{\rm tot} = \int _{E_1} ^\infty dE \, E \,Q(E) \,,
 \label{eq:Etot}
\end{equation}
where we fix $E_1 = 0.1$ GeV. The normalization of the spectrum in Eq.~\ref{eq:Q_E} can be constrained from available catalog quantities for single SNRs and PWNe, 
or by using average population characteristics for the smooth component. 
\\
The normalization for a single PWN is obtained assuming that a fraction $\eta$ of the total spin-down energy $W_0$ emitted by the pulsar is released in form of $e^\pm$ pairs, {\it i.e.}:
\begin{equation}
E_{\rm tot,PWN} = \eta W_0. 
\label{eq:EtotPWN}
\end{equation}
The value of $W_0$ can be computed starting from the age of the pulsar $t^{\star}$, 
the typical pulsar decay time $\tau_0$ and the spin-down luminosity $\dot{E}$:
\begin{equation}
 W_0 = \tau_0 \dot{E} \left( 1+ \frac{t^{\star}}{\tau_0} \right)^2\,.
 \label{eq:W0PWN}
\end{equation}
The spin-down luminosity $\dot{E}$, the \textit{observed} age $t_{\rm obs}$ (where $t^{\star} = t_{\rm obs} + d/c$ is the \textit{actual} age) and distance for each pulsar are taken from \cite{2005AJ....129.1993M}, 
while $\tau_0=10$ kyr. In the following of our analysis, the free parameter associated with the PWN spetrum normalization is  $\eta$. 
\\
The normalization of a single SNR  spectrum can be linked to its radio brightness $B_r(\nu)$ at a frequency $\nu$. 
Assuming that $B_r(\nu)$ is due to synchrotron emission of the surrounding electrons in the SNR magnetic field $B$, 
we obtain (see \cite{2010A&A...524A..51D}):
\begin{equation}\label{eq:snrQ0}
Q_{0, \rm SNR} = 1.2 \cdot 10^{47} \text{GeV}^{-1} (0.79)^{\gamma_{\rm SNR} }
 \left[ \frac{d}{\text{kpc}}\right]^2  \\
  \left[\frac{\nu}{\text{GHz}} \right]^{\frac{(\gamma_{\rm SNR} -1)}{2}} 
 \left[ \frac{B}{100 \mu\text{G}} \right]^{-\frac{(\gamma_{\rm SNR} +1)}{2}} 
 \left[ \frac{B^\nu _r}{\text{Jy}}\right]
\end{equation}
where $d$ is the source distance and the electron energy spectrum index $\gamma_{\rm SNR}$ in Eq.~\ref{eq:snrQ0} is related to the radio  index $\alpha_r = (\gamma_{\rm SNR}  -1)/2$.
For each source, we take the value of $B_r(\nu)$ and  $\gamma_{\rm SNR} $ from the Green catalog \citep{Green:2014cea}. The remaining parameters (distance, age, magnetic field) are taken from other literature results (see Table 1 in \cite{2014JCAP...04..006D}). 
In our analysis, a special focus is deserved to the Vela SNR. 
Depending on $R_{\rm cut}$, the contribution of the \textit{near} SNRs is considered as the sum of all the single SNRs in the Green catalog plus, separately,  the electrons arriving from Vela. Each \textit{near} SNR is normalized according to Eq.~\ref{eq:snrQ0}. We only consider a free normalization parameter $N_{\rm near}$, which is 
the rescaling factor of the normalization of sum of all the \textit{near} SNRs (Vela excluded, indeed). 
As for the Vela SNR, we let its normalization free to vary separately. We express it through the connection with its magnetic field $B_{\rm Vela}$ by means of Eq.~\ref{eq:snrQ0}.
\\
The normalization of  the electron spectrum for the smooth, {\it far}  SNR distribution can be connected to the average properties of the entire Galactic population 
 by means of Eq.~\ref{eq:Etot}, as described in \cite{2010A&A...524A..51D}. 
We will express the free parameter for the energy spectrum normalization in terms of the total energy carried by electrons $E_{\rm tot,SNR}$.
As for the average Galactic SN explosion rate, we fix $\Gamma_*=$1/century.
The values we  obtain are checked to lie in a range of plausibility as determined in \cite{2010A&A...524A..51D}.

The solution to the diffusion equation for a smooth population ({\it i.e.} of SNRs) can be studied in terms of the 
 \textit{halo function}, which describes the probability for an electron to reach the Earth position $\mathbf{x_{\odot}}$ from its source located at 
$\mathbf{x_s}$:
\begin{equation}
\label{eq:hf}
\Upsilon(\lambda) = \int d^3 \mathbf{x_s} \rho(\mathbf{x_s}) \mathcal{G}_{\lambda}(\lambda, x_{\odot} \leftarrow \mathbf{x_s}),
\end{equation}
for a given normalized spatial distribution of sources $\rho(\mathbf{x_s})$. 
The  propagation scale length $\lambda$  is given by:
\begin{equation}
\label{eq:lambda}
 \lambda^2= \lambda^2 (E, E_s) \equiv 4\int _E ^{E_s} dE' \frac{K(E')}{b(E')}. 
\end{equation} 
with the diffusion coefficient $K(E)$ and the energy-loss rate $b(E)$ previously defined.
The  integral in Eq.~\ref{eq:hf} is performed over all the diffusive volume extending up to $L$.  The Green function of the transport equation can be cast as:
\begin{equation}
\label{eq:green}
 \mathcal{G}_{\lambda}(\lambda, \mathbf{x_{\odot}} \leftarrow \mathbf{x_s}) \equiv b(E)\mathcal{G}(E, \mathbf{x_{\odot}} \leftarrow E_s, \mathbf{x_s})\,.
\end{equation}
 \\
The flux of electron $\Phi$ at the Earth or, equivalently the number density $\psi$ ($\Phi=v /4\pi \;\psi$), may be generically computed from the Green
function $\mathcal{G}_{\lambda}(\lambda, \mathbf{x_{\odot}} \leftarrow \mathbf{x_s})$ and the source term ${Q}$  after an integration  
over time, energy and the spatial extent of the diffusion zone: 
\begin{equation}
\label{eq:flux_def}
\Phi (E) = \frac{\beta \, c}{4\,\pi} 
 \int dt_sdE_s d^3\mathbf{x}_s \, 
\mathcal{G}_{\lambda}(\lambda, \mathbf{x_{\odot}} \leftarrow \mathbf{x_s}) 
{ Q}(t_s,E_s,\mathbf{x}_s)
\end{equation}
where the subscript $s$ refers to quantities at source, and  $t_s$ is the source age.  
As for the Green function, we will adopt the approximation in which only the vertical boundaries of the diffusion halo are considered, since the radial boundary has been demonstrated to be irrelevant while $r-r_{\odot}\gtrsim L $ \citep{2010A&A...524A..51D}. 
Thus, the spatial dependence in Eq.~\ref{eq:green} is separated in a radial term and in a vertical term as $\mathcal{G}_{\lambda} = \mathcal{G}_r \times \mathcal{G}_z$.
For the radial term we use the Green function for an infinite 2D space. 
The vertical boundary is accounted for by expanding $\mathcal{G}_z$ with the image method \citep{Baltz:1998xv} or the Helmotz eigen-functions \citep{Lavalle:2006vb},
depending on the propagation scale length.
\\
A steady state solution is adopted for the secondary production, 
since this gives a continuos injection of electrons and positrons in the ISM. 
This is the case also for smooth distributions of sources, described by average parameters as the SN explosion rate. 
Conversely, the solution of the time-dependent transport equation is required for the flux from a single source of electrons and can be found in several literature works (see  \cite{Mlyshev:2009twa, 2010A&A...524A..51D}), 
to which we address the reader for further details. 
Because of our focus on single sources, we remind briefly the explicit expression in the burst-like approximation, 
namely $Q(E, \mathbf{x}, t)= Q(E) \delta(t) \delta(\mathbf{x})$:
 \begin{equation}\label{eq:singlesourcesolution}
  \psi(\mathbf{x_{\odot}}, E,t) = \frac{b(E_s)}{b(E)} \frac{1}{(\pi \lambda^2)^{\frac{3}{2}}} \exp\left({-\frac{|\mathbf{x_{\odot}} -\mathbf{x_{s}} |^2}{ \lambda^2}}\right)Q(E_s)
\end{equation}
where $E_s$ is the initial energy of electrons that cool down to $E$ in a \textit{loss time}
\begin{equation}
 \Delta \tau (E, E_s) \equiv \int_E ^{E_s} \frac{dE'}{b(E')} = t-t_s .
\end{equation}
The halo function defined in Eq.~\ref{eq:hf}  represents the probability of an electron (or positron) to reach the Earth position, given its propagation 
scale and the spatial distribution of its sources. To explore the role of the radial cut on the smooth distribution of sources, as well 
of the propagation model, we analyze  $\Upsilon(\lambda)$ as a function of the propagation scale $\lambda$. 
\begin{figure}[t] 
  \centering\includegraphics[width=0.6\textwidth]{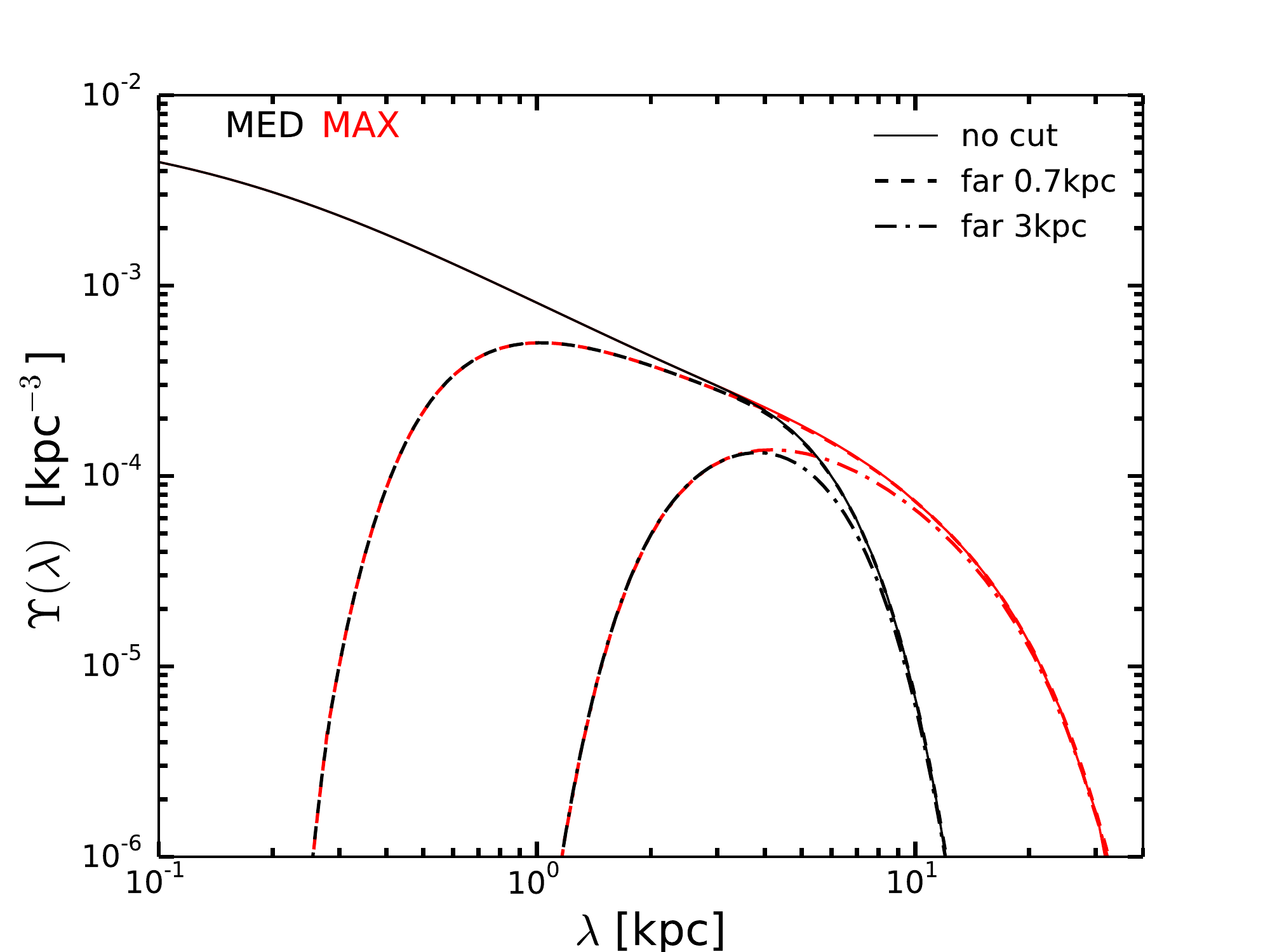}
   \caption{Halo function for the G15 smooth distribution of 
SNRs for different values of the radial cut around the Earth position, and for 
the MED and MAX propagation models. Solid black (red) line corresponds to a 
smooth distribution with no cut and MED (MAX) propagation model, with $L=4$ 
($15$) kpc. Dashed (dot-dashed) lines correspond to a SNR  
distribution beyond a radial cut of $0.7$~kpc ($3$~kpc).}  \label{fig:halofunction}
\end{figure}
In Fig.~\ref{fig:halofunction} we show  $\Upsilon(\lambda)$  as a function of  $\lambda$, for two
different values of $R_{\rm cut}$ and the two propagation benchmarks.
The two solid lines correspond to the case of $R_{\rm cut}=0$, namely when all the Galactic SNRs are shaped according to Eq.~\ref{eq:rho}. 
The probability for an electron to reach the Earth decreases shortly to zero for propagation length $\lambda  \gtrsim L$  ($L$=4 (15) kpc for the MED (MAX) model).
The halo function of the {\it far} SNR population has a similar trend at high $\lambda$ but it drops to zero for propagation 
lengths roughly below the size of the radial cut. 
The halo function for the MED and MAX models is undistinguishable for $\lambda \lsim 4$ kpc, 
while for $\lambda \gsim 4$~kpc we find that $\Upsilon(\lambda)$  is greater for higher diffusive haloes. Electrons have greater chances to arrive at the Earth
if the diffusion zone is wider.
The action of a radial cut of the smooth SNR distribution has the effect of depleting the injection of electrons from near SNRs. 
The halo function for the particles arriving from $R>R_{\rm cut}$ is smaller for larger cut radii, meaning that it gets less probable to reach the Earth 
 for electrons produced farther. This is clearly visible in Fig.~\ref{fig:halofunction} where, for a fixed propagation scheme, 
 $ \Upsilon(\lambda)$ for the {\it far} electrons is  small, and dropping to zero for propagation lengths roughly smaller than the radial cut. 
\begin{figure*}[t]
\centering
\flushleft
\includegraphics[width=1.05\textwidth]{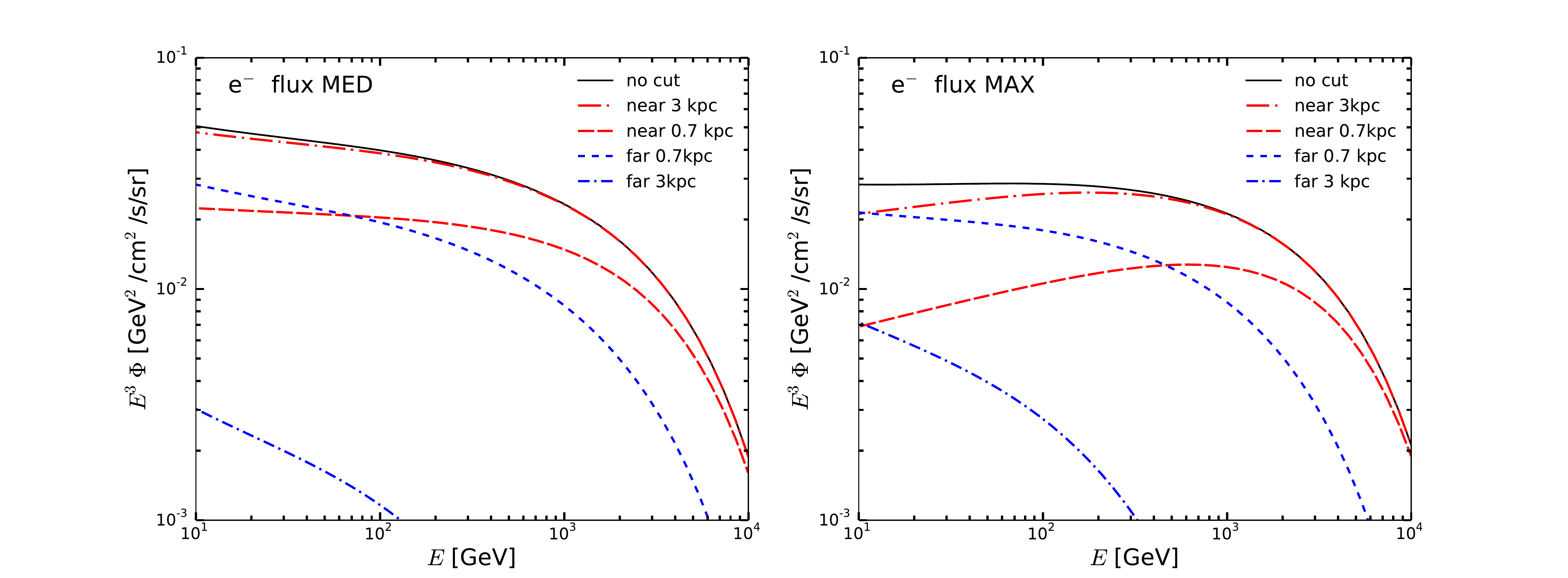}\ 
\caption{Electron fluxes for different SNR smooth populations with a G15 radial 
distribution. Black solid line corresponds to a SNR distribution 
with no cuts; red long dot-dashed (long dashed) line corresponds to a { \it near} SNR 
distribution of radius $R_{\rm cut}= 3$ kpc ($0.7$ kpc) around the Earth; blue 
dashed (dot-dashed) line corresponds to a {\it far} SNR distribution from which the 
smooth component up to a radius $R_{\rm cut}= 0.7$ kpc ($3$ kpc) has been cut. 
All fluxes are obtained for a SNR energy 
spectrum with $E_c=5$~TeV, $\gamma=2.3$, $E_{\rm tot,SNR}=10^{49}$ erg, left (right) panel
for the MED (MAX) propagation model.} \label{fig:smoothfluxes}
\end{figure*}

The effect of different $R_{\rm cut}$ values and propagation models on the flux of electrons at Earth are shown in Fig.~\ref{fig:smoothfluxes}. 
We plot the electron flux from both the {\it far} and the {\it near} SNR smooth distributions, 
for $R_{\rm cut}$ =  0.7 and 3 kpc and for no cut, and for the MED and MAX models. 
The energy cut-off in Eq.~\ref{eq:Q_E} is fixed to $E_c$ = 5~TeV and the spectral index is $\gamma=2.3$. 
We have verified  that for a given value of $R_{\rm cut}$, the  {\it far} and  {\it near} components sum up exactly to the total flux with no cuts.
The flux from a smooth Galactic SNR population following the radial profile G15 is shown by solid lines. 
It is a function decreasing with energy slightly stronger than $E^{-3}$, to drop exponentially to zero when approaching the cut-off energy. 
The flux from sources inside $R_{\rm cut}$ =3~kpc is very close to the flux from a smooth population all over the Galaxy (no cut case). 
The difference is significant only for the MAX case (large diffusive haloes), 
where there is a reduction in the near flux at energies below few hundreds GeV. 
The flux from sources located inside $R_{\rm cut}$ = 0.7~kpc is instead smaller
than the flux from a smooth population all over the Galaxy,
with  the high energy tail asymptotically converging to the no cut case. 
The electrons coming from the {\it far} population show a trend similar to the whole smooth population, 
but decreased by a rough factor of two. 
Comparing the {\it far} and {\it near} fluxes, we find, as expected, that the most energetic electrons come from the closest sources, as
firstly  noted in \cite{Aharonian:1995zz}.
The flux of electrons coming from $R>3$ kpc is one order of magnitude smaller than for the sources at $R>$0.7 kpc in the MED case, 
a factor of two in the MAX case. 
For higher diffusive haloes, electrons from far sources have greater chances to reach our detectors. 
Our results have been obtained for a G15 source distribution. 
We have checked that the L04 radial profile gives fluxes systematically higher by a rough 10\%, 
since it  predicts more sources near the Earth position, as shown in Fig.~\ref{fig:rho}. 

 \begin{figure}[th]
  \centering
  \includegraphics[width=0.8\textwidth]{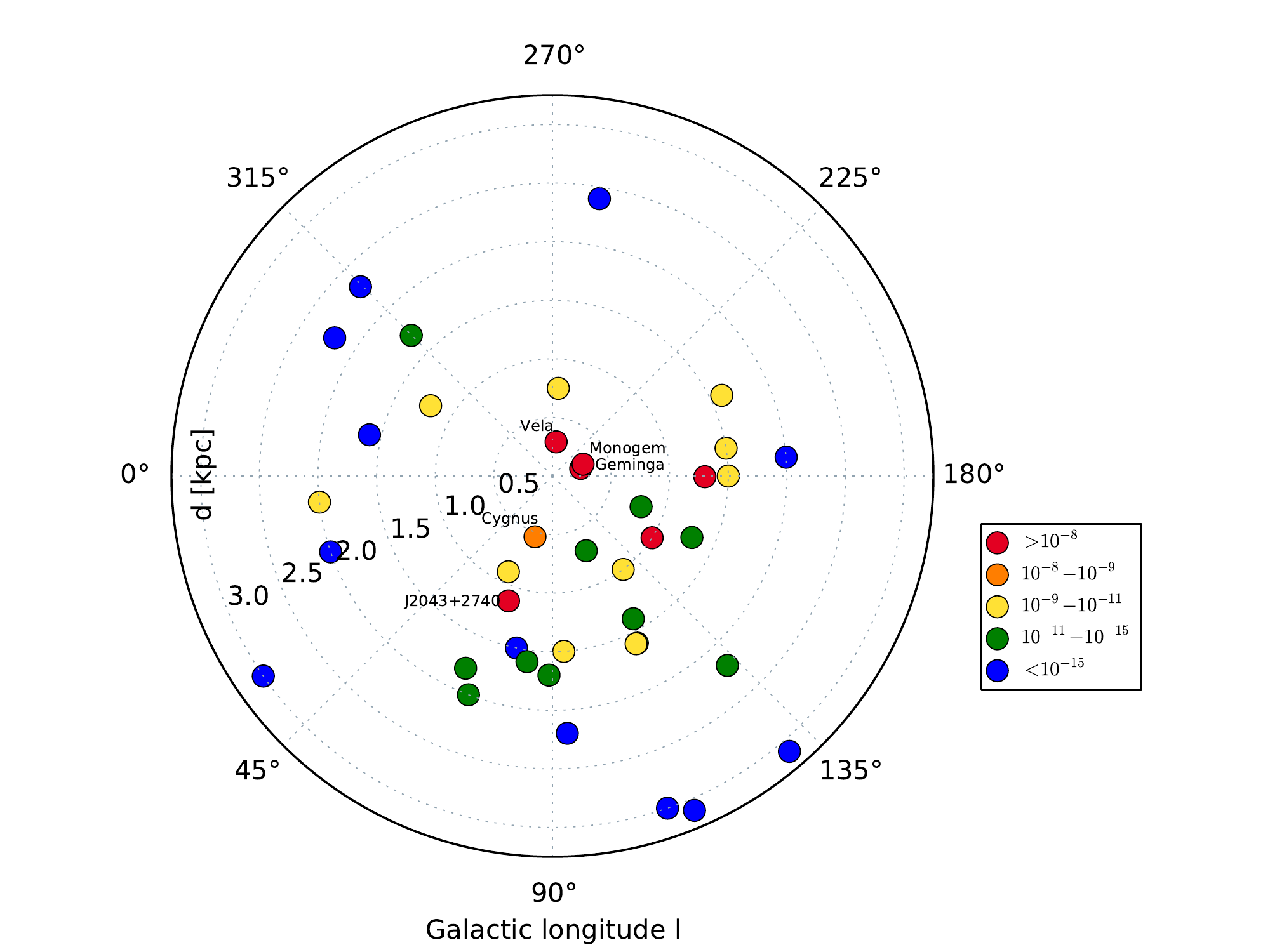}\
   \caption{This figure shows the sample of single SNRs and PSRs in terms of their Galactic longitude $l$ [deg] and  distance to the Earth [kpc] (located at the center of the circle). 
   The color scale of the dots quantifies the electron flux  integrated from $50$~GeV to $5$~TeV, in units of ${\rm ( cm ^{2} \; s \; sr)^{-1}}$.}  
   \label{fig:sourceposition}
 \end{figure} 
In Fig.~\ref{fig:sourceposition} we plot our sample of {\it near} SNRs together with the most powerful PWNe identified in \cite{2014JCAP...04..006D}, 
projected on the Galactic plane. The SNR and PSR characteristics are taken  from the Green \citep{Green:2014cea} and the ATNF \citep{2005AJ....129.1993M} catalogs, respectively. 
The source position is identified  by the Galactic longitude ($l$, deg) and distance ($d$, kpc) to the Earth. 
The color scale reflects the intensity of the integrated electron flux at the Earth from $E=50$~GeV up to $E=5$~TeV, which is an important observable when computing the anisotropy from a single source. 
Fluxes are computed by means of Eq.~\ref{eq:singlesourcesolution}, using available catalog parameters and the MAX propagation model. 
The spectral index for PWNe is fixed to $\gamma_{\rm PWN} = 1.7$, while the conversion to electron efficiency is $\eta = 0.07$. 
Vela (red dot) and Cygnus Loop (orange dot) are the  SNRs with the highest integrated flux.  
The most powerful PWNe Monogem, Geminga and J2043+2740,  give an integrated electron flux above $10^8$ ${\rm (cm ^{2} \; s \; sr)^{-1}}$. 
We report in Table~\ref{tab:integratedflux} the catalog characteristics of the most powerful  SNRs and PSRs in Fig.~\ref{fig:sourceposition}, along with their 
 electron flux $\phi_{50}$ integrated from $50$~GeV to $5$~TeV.
\begin{table}
\caption{Properties for the closest and most powerful 
SNRs and PSRs shown in Fig.~\ref{fig:sourceposition}. 
The first column indicates the source name in the Green SNR, (first two lines) and in the ATNF PWN, (from third line down) catalog; 
the second column shows other used names for some of the sources; the third and fourth columns report the distance (in kpc) and the age of the source (in kyr), respectively, while the last column reports the integrated electron flux $\phi_{50}$ from $50$~GeV to $5$~TeV in  units of ${\rm ( cm ^{2} \; s \; sr)^{-1}}$.}
\label{tab:integratedflux}
 \begin{center}
 \begin{tabular}{l l c c c}
Source & other name        & d [kpc]        & $t_{\rm obs}$ [kyr]         & $\phi_{50} [{\rm (cm ^{2} \; s \; sr)^{-1}}$]  \\ \toprule
 263.9-3.3                &Vela YZ          & 0.293         & 11.4                &  1.3 $\cdot$ 
10$^{-7}$ \\ 
 74.0-8.5                 & Cygnus Loop          &  0.54                & 10                & 3.2 $\cdot$  
10$^{-9}$ \\  \midrule
 J0633+1746         &Geminga          & 0.25                 &343                 & 1.7 $\cdot$  
10$^{-8}$\\ 
 J2043+2740        &                  &1.13                  & 1200                &3.5 $\cdot$  10$^{-8}$  
\\ 
 B0355+54        &                  &1                 &564                 &1.6 $\cdot$  10$^{-8}$  \\ 
 B0656+14        & Monogem        & 0.28          &111                 & 1.1 $\cdot$ 10$^{-8}$  \\ 
 J0538+2817        &                  &1.3                  & 618                & 1.3 $\cdot$ 10$^{-8}$ \\ 
\bottomrule
\end{tabular}
\end{center}
\end{table}

\medskip
\section{Anisotropy in the diffusion model}
\label{sec:thframe}

A detailed study of the anisotropy in the CR flux should be based on a full spherical harmonic analysis. 
In that case, the flux intensity would get expanded  in multipoles as a function of direction: $I(\theta, \phi) = \sum A_{lm} Y_{l}^{m}(\theta, \phi)$, 
where $ Y_{l}^{m}(\theta, \phi)$ are the spherical harmonics and $A_{lm}$ are the constant coefficients. However, under the hypothesis of one (or few) dominant nearby source, we expect the dipole term to dominate the multipole expansion. 
In this case, the anisotropy of CRs can be defined as
 \begin{equation}
 \Delta = \frac{I_{max}- I_{min}}{I_{max}+ I_{min}}\,,
\end{equation}
where $I_{max}$ and $I_{min}$ are, respectively, the maximum and minimum values of the CR intensity. This expression can be computed in a diffusive propagation regime, as early derived  by \cite{1964Ginzburg}:
\begin{equation}\label{eq:genericdipole}
 \Delta = \frac{3 K}{c} \left| \frac{\nabla \psi}{\psi} \right|\,.
\end{equation}
An explicit expression for the dipole anisotropy can be derived from the solution to the number density $\psi$ at the Earth, as got explicit  in Eq.~\ref{eq:singlesourcesolution}. As an example, the electron plus positron anisotropy from a single electron source $s$ is given by:
\begin{equation}
\label{eq:eleposdipole}
  \Delta(E)_{e^+ + e^-} = \frac{3 K(E)}{ c} \frac{2 d_{s}}{\lambda^2(E, E_{s})} \frac{\psi_{e^+ + e^-}^{s}(E)}{\psi_{e^+ + e^-}^{tot}(E)},
\end{equation}
where $d_s$ is the distance to the source, $\lambda(E, E_{s})$ is the propagation scale defined in Eq.~\ref{eq:lambda}, $\psi_{e^+ + e^-}^{s}(E)$ is the $e^+ + e^-$ number density produced by the source $s$, and $\psi_{e^+ + e^-}^{tot}(E)$ is the total $e^+ + e^-$ number density obtained from the contributions of all the sources, both from isotropic smooth populations and from directional single sources.

More generally, if we are dealing with a collection of electron and/or positron sources, following the early work of \cite{1971ApL.....9..169S}, the total dipole anisotropy may be computed as:
\begin{equation}
\label{eq:dipolesources}
 \Delta(n_{max}, E)= \frac{1}{\psi^{tot}(E)} \cdot \sum_i \frac{\mathbf{r}_i\cdot \mathbf{n}_{max}}{||\mathbf{r}_i||}\cdot \psi_i(E)\, \Delta_i(E).
\end{equation}
Here $\psi_i(E)$ is the  number density of electron and/or positron emitted from each source $i$, $\mathbf{r}_i$ is the 
 source position in the sky and 
$n_{max}$ is the direction of the maximum flux intensity. The term $\psi^{tot}(E)=\sum_i \psi_i(E)$ is the total (electron and/or positron) number density and includes the contribution from the discrete as well as all the isotropic sources. 
The anisotropy from each single source is given by $\Delta_i = \frac{3 K(E)}{c} \frac{|\nabla \psi_i(E) |}{\psi_i(E)}$, where the gradient is performed with respect to each source position. 

In the next Sections we will provide results both in the approximation in which a single dominant source is responsible for  the whole dipole anisotropy, as well as for a collection of close, discrete sources.

\medskip
\section{Fit to AMS-02 fluxes: method and analysis}
\label{sec:fluxes}
  
One of the main aims of our analysis is to provide predictions for the anisotropies in the lepton sector only for those models 
compatible with the observations of electron and positron fluxes. 
In particular, in \cite{2014JCAP...04..006D, 2016JCAP...05..031D} it is shown that the AMS-02 data on the lepton fluxes can be consistently described in terms of the contributions from PWNe and  SNRs and a secondary component produced by the spallation of cosmic proton and helium in the ISM. 
Here, we fit different models to the AMS-02 data on the $e^+$ \citep{2014PhRvL.113l1102A} and $e^+ + e^-$ \citep{2014PhRvL.113v1102A} fluxes. 
As for the secondary production of  $e^+$ and $e^-$, it is modeled as \cite{Delahaye:2008ua} and \cite{2014JCAP...04..006D} (to which we refer for any detail),  
we let it to shift by a free normalization $\tilde{q}_{sec}$. 
The  SNRs and PWNe are modeled as described in the previous Sections. 
The SNR contribution for a smooth distribution of sources is computed with a G15 radial distribution. 
If not differently stated, the propagation parameters are set to the MAX configuration.
We fit data points for $E>$ 10 GeV, in order to avoid biases from the solar modulation of the fluxes. Nevertheless, we let the flux be reshaped by the solar wind  
according to the force field model,  with the modulation potential $\phi_F$ as a free parameter. 
 We have worked within four different schemes, mostly different for the treatment of local source contributions, 
which are listed and briefly  outlined below. 

\medskip 

{\textit{\textbf{Case 1.} - Fixing Vela SNR parameters}}
\\
The first analysis we undertake is meant to emphasize the role of the Vela SNR  shaping the high energy AMS-02 data and, consequently, the anisotropy level. 
We fit the $e^+$ and $e^+ +e^-$ AMS-02 data for  $R_{\rm cut}$ = 0, 0.7 and 3 kpc. 
The flux of $e^+$ and $e^-$ from  PWNe is computed for all the ATNF catalog sources as  outlined in Sect.~\ref{sec:generics}. 
The free parameters in the smooth SNR modeling are the spectral index $\gamma_{\rm SNR}$ and the total energy emitted in electrons $E_{\rm tot, SNR}$.
When $R_{\rm cut}\neq 0$, the electron flux from the \textit{near} SNRs is computed as described in Sec.~\ref{sec:generics} (in particular Eqs.~\ref{eq:singlesourcesolution} and \ref{eq:snrQ0}), using available data on the radio spectrum, distance and age for each source. 
The contribution of all the SNRs within $R\leq R_{\rm cut}$ except Vela is summed up to a total \textit{near} SNR flux. The normalization $N_{\rm near}$ of this component 
is a free parameter of the fit. As a matter of fact, this sum is dominated by Cygnus SNR, which overclasses the contribution from any other near SNR by 
at least two orders of magnitude. Therefore, the parameter $N_{\rm near}$ can be effectively associated to the Cygnus magnetic field. 
Exception is deserved to the brightest local SNR Vela  (see Fig.~\ref{fig:sourceposition}), 
for which  the value of the magnetic field is treated as a free parameter. 
The radio emission from the Vela SNR environment has been studied in  detail in \cite{2001A&A...372..636A}. 
From spectral index and morphology arguments, the region labeled Vela X is thought to be connected to a PWN from the young Vela pulsar.
We  use therefore the radio brightness associated to the Vela YZ radio region only, which is reported to be
 $ B_r= (588+547)$ Jy $= 1135 $Jy at $ 980 $MH \citep{2001A&A...372..636A}.
In the same study, a radio index of $\alpha_Y = 0.70 \pm 0.10$ and $\alpha_Z = 0.81 \pm 0.16$ is found, and a mean spectral 
index of $\gamma_{\rm YZ}= 2.5 \pm 0.3$ for the Vela YZ region can be inferred. 
We fix here the Vela spectral index to 2.5.
Its age is set to $t_{\rm obs}= 11.4$~kyr by \cite{1970ApJ...159L..35R}. 
The value of the magnetic field surrounding known SNRs is quite uncertainty and still a matter of debate, 
and it is connected to the leptonic or hadronic origin of the gamma ray emission from young SNRs (for a review see \cite{Reynolds:2011nk}).
We vary the Vela magnetic field, $B_{\rm Vela}$, in the range $[1, 100]\,\mu$G.
The radio properties of the Cygnus SNR are taken from \cite{2006A&A...447..937S}.
\\
{\it Case 1} has therefore the following free parameters:  $\phi_F$, $\tilde{q}_{sec}$, $\gamma_{\rm PWN}$, $\eta$, $N_{\rm near}$, $B_{\rm Vela}$, $\gamma_{\rm SNR}$, $E_{\rm tot,SNR}$. 
\\
The results of the fits on $e^++e^-$ and $e^+$ AMS-02 data are in general good for $R_{\rm cut}$ = 0, 0.7 and 3 kpc, 
both for the MED and MAX propagation parameters. 
All the fits have a reduced chi-square $\chi^2/d.o.f.$ of the order or smaller than 1.
The MAX scenario is slightly better than the MED one for $R_{\rm cut}\neq0$ and the model with $R_{\rm cut}=0.7$ is preferred with respect to the other cases (the chi-square for the MAX model for the cuts $R_{\rm cut}$ = 0, 0.7 and 3 kpc are 53, 38 and 41 respectively).
Vela has indeed an important role on shaping the flux at few hundreds GeV. We will inspect more closely its role in the following.
We checked  the effect of a lower energy cutoff for the Vela SNR, working with $E_c=2$~TeV. We found that the best fit parameters for the SNR smooth, the secondary and the PWNe components are left almost unchanged, since the effect of the cutoff  is effective at energies above the AMS-02 data. 
For example, for $E_c=2$ TeV the Vela flux (multiplied by $E^3$) at 5 TeV is a factor of $\sim 4$ lower than the flux obtained with  $E_c=5$ TeV. 

\medskip
  
\begin{figure*}[t]
\includegraphics[width=0.52\textwidth]{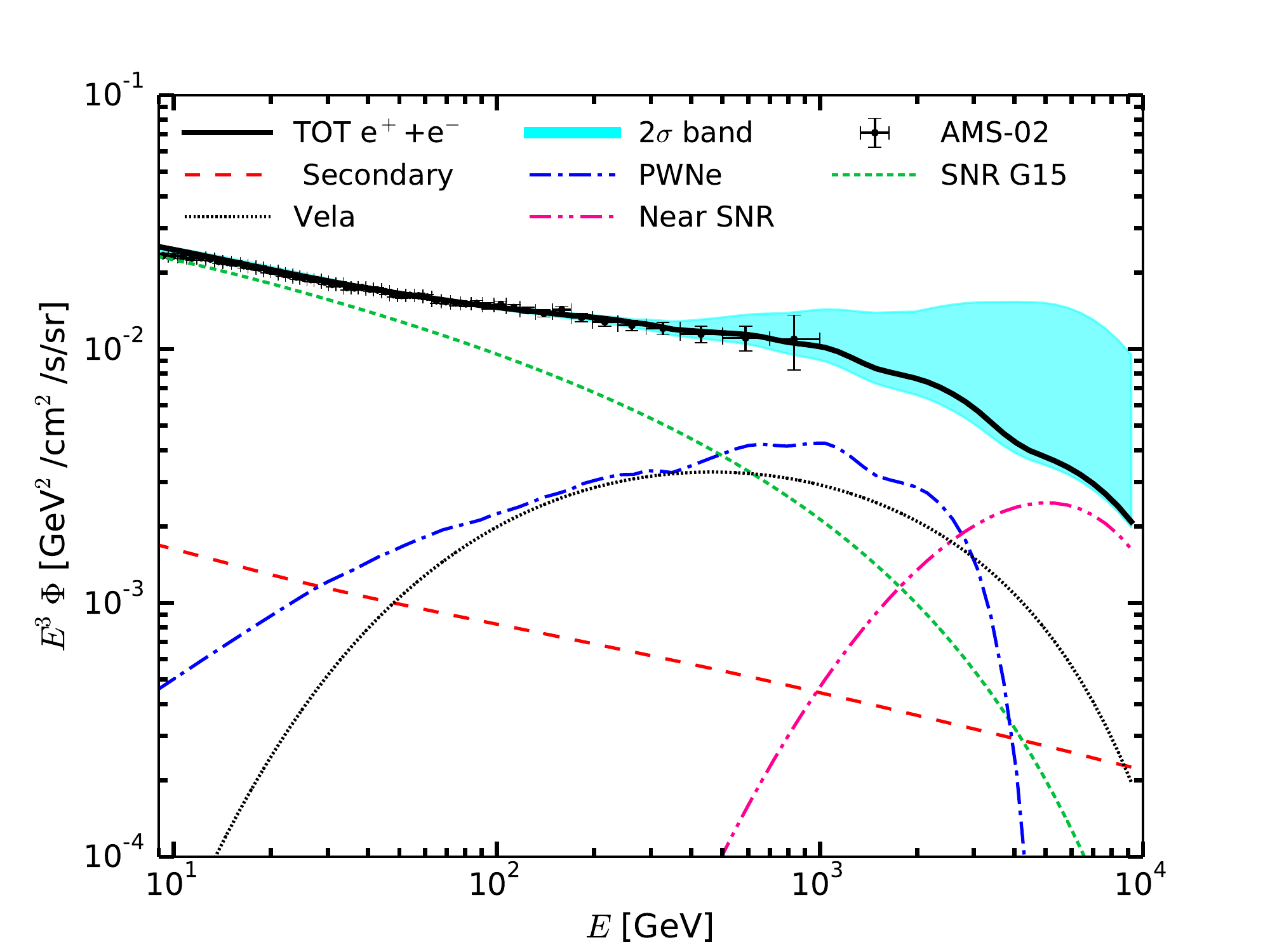}
\includegraphics[width=0.52\textwidth]{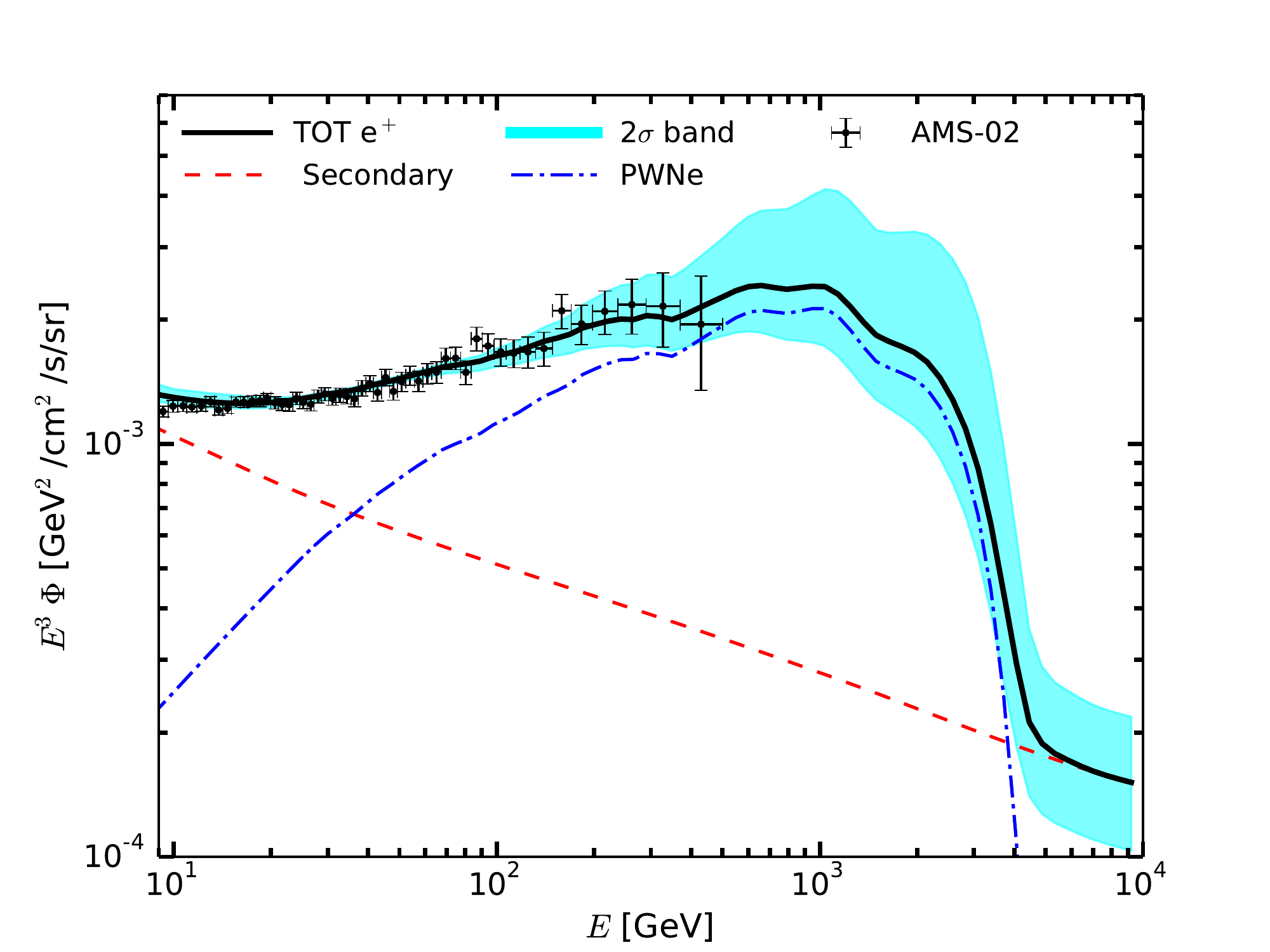}
\caption{Fit to $e^++e^-$ (left panel) and $e^+$  (right panel) AMS-02 data \citep{2014PhRvL.113v1102A, 2014PhRvL.113l1102A} 
with MAX propagation model and $R_{\rm cut}=0.7$~kpc for 
\textit{Case 2} (see text for details).
All the components for the best fit are displayed together with the 2$\sigma$ uncertainty band on the total flux. 
Line coding as follows: 
solid black, sum of all the components in the plot; 
red dashed,  secondary $e^+$ and $e^-$; 
 blue dash-dotted: $e^+$ and $e^-$ from all ATNF PWNe;
 green dotted: $e^-$ from far  SNR; 
 black dotted: $e^-$  from Vela SNR; 
magenta double dash-dotted: $e^-$  from all other near ($R \leq 0.7$~kpc) SNRs. 
The left (right) panel shows the $e^++e^-$ ($e^+$) flux.} 
\label{fig:cut07kpc_case2}
\end{figure*}

{\textit{\textbf{Case 2}. \it - Insights on Vela SNR parameters.}} 
\\
Since the Vela SNR is the most intense local source we inspect its distinctive parameters with more details. 
As well as its magnetic field ({\it Case 1}), we also admit variations in its distance and spectral index, whose measurements 
deal with non negligible uncertainties. 
The Vela distance is found to be $d_{Vela} =0.293^{+0.019} _{-0.017}$~kpc in \cite{Dodson:2003ai} with proper motion and parallax measurements for the Vela pulsar, 
while a lower value is suggested by high resolution absorption line spectra measurements of stars in the Vela SNR direction $d_{Vela} =0.250 \pm0.030$~kpc  \citep{Cha:1999pn}. 
To account for these uncertainties, we explore $d_{Vela} = [0.22, 0.32]$~kpc. 
As for the  spectral index of the remnant, together with the magnetic field, it affects the shape of the energy spectrum and the normalization of the flux. 
It spans here an interval of $\gamma_{\rm Vela} = [2.0, 2.8]$ (see discussion at {\it Case 1}). 
As for {\it Case 1}, the remnant magnetic field  is moved in the range $B_{\rm Vela} = [1, 100]\,\mu$G.
\\
The free parameters for this second analysis are thus  $\phi_F$, $\tilde{q}_{sec}$, $\gamma_{\rm PWN}$, $\eta$, $N_{\rm near}$, $B_{\rm Vela}$,  $d_{\rm Vela}$, $\gamma_{\rm Vela}$, $\gamma_{\rm SNR}$ and $E_{\rm tot,SNR}$. 
In Fig.~\ref{fig:cut07kpc_case2} we plot the results for the fit to $e^++e^-$ (left panel) and $e^+$ (right panel) AMS-02 data, 
for $R_{\rm cut}$=0.7 kpc and MAX propagation model for this \textit{Case 2}. 
The best fit for the total flux is drawn together with its 2$\sigma$ uncertainty band. The minimum $\chi^2$ is 32 for 98 data points. 
The fit to both the $e^+$ and $e^++e^-$ data is remarkably good, and its uncertainty spreads accordingly to the experimental error bars, 
which are very small up to about 200 GeV. The uncertainty band has been extrapolated beyond the experimental maximal energy of about 1 TeV, 
showing its spread up to a factor of three at 5 TeV. A similar plot is obtained for {\it Case 1}, if it were not for a narrower extrapolated uncertainty band.
In Tab.~\ref{tab:case2velabf} the values of the best fit parameters for \textit{Case 2} are shown together with their errors.
The parameters for the SNR smooth population and the ATNF PWNe are found to lie intervals similar to \textit{Case 1} and consistent with previous results \cite{2014JCAP...04..006D,2016JCAP...05..031D}.
As for the parameters most relevant to the following of our analysis,
the best fit for the Vela distance is found to be very near to the measurement of \cite{Dodson:2003ai}, 
while the spectral index points to the higher permitted value of 2.8. 
This result can be hardly argued  with the general modeling of diffusive shock acceleration in SNRs. 
However, we find numerous configurations included in the $2\sigma$ band with lower values for $\gamma_{\rm Vela}$ of 2.4-2.6, while the rest of the parameters keeping similar values to the ones reported in Tab.~\ref{tab:case2velabf}.

 \begin{table}
 \centering
 \caption{Best fit parameters to AMS-02 $e^++e^-$ and $e^+$ flux data for the model described by \textit{Case 2}.} 
 \begin{tabular} {*{2}{l|l}} \toprule
 $\phi_F$ & $0.36$ GV\\ 
 $\tilde{q}_{sec}$ & $2.10 \pm 0.08$ \\
 $\gamma_{PWN}$& $1.85 \pm 0.03$  \\
 $\eta$& $0.065\pm0.004$ \\  
 $N_{\rm near}$& $0.35 \pm 0.03$ \\
 $B_{Vela}$&$(3.1\pm0.3)$ $\mu$G\\
 $d_{Vela}$&$0.29\pm0.04$~kpc\\ 
 $\gamma_{Vela}$&$2.80$\\ 
 $\gamma_{ \rm SNR}$& $2.65\pm0.03$ \\
 $E_{\rm tot,SNR}$&$(3.50 \pm0.05)10^{49}$~erg\\ 
 $\chi^2/$d.o.f &32/89 \\ \bottomrule
 \label{tab:case2velabf}
  \end{tabular}  
 \end{table}

\medskip

{\textit{\textbf{Case 3.} - An unknown close SNR.}}
\\
In the previous analysis we have considered  only close SNRs with a detected electromagnetic counterpart. 
They account for a population of relatively young remnants, with ages at most of 50-100 kyr. 
However, due to the diffusion time, electrons accelerated by older SNRs may contribute to the flux at Earth while being
 no longer visible in any electromagnetic band. 
Here we analyze the AMS-02 data in the hypotheses that the only SNR injecting electrons in the local Galaxy, 
namely for  $R \leq R_{\rm cut}$=0.7 kpc, is a source no longer detectable in radio or any other electromagnetic frequency. 
We let the age of this unknown SNR vary in the range 50 kyr - 10 Myr,  the distance between 0.1 and 0.7 kpc, the injection 
spectral index $\gamma$ between 2.0 and 2.6. Its normalization is a free parameter, with the constraint not to overtake
$E_{\rm tot} = 10^{49}$ erg \citep{2010A&A...524A..51D}. 
This object might  be connected, for instance, to the Local Bubble, 
which in some models is explained as the result of a single or multiple supernovae exploding $10^5-10^6$~yr ago in proximity of the solar system  \citep{2001ApJS..134..283S,Helmut_LB_2011}.
\\
\begin{figure*}
\includegraphics[width=0.52\textwidth]{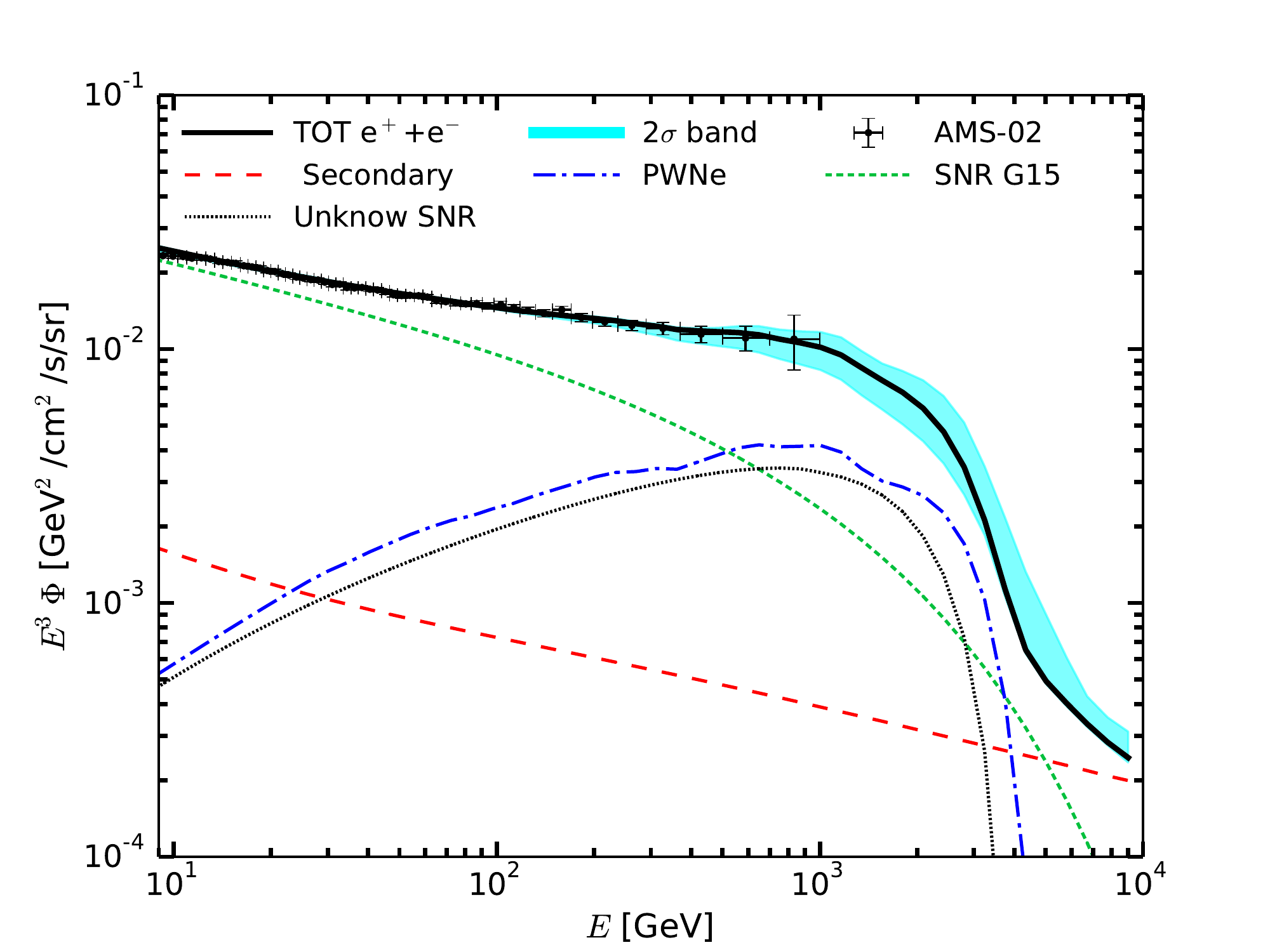}
\includegraphics[width=0.52\textwidth]{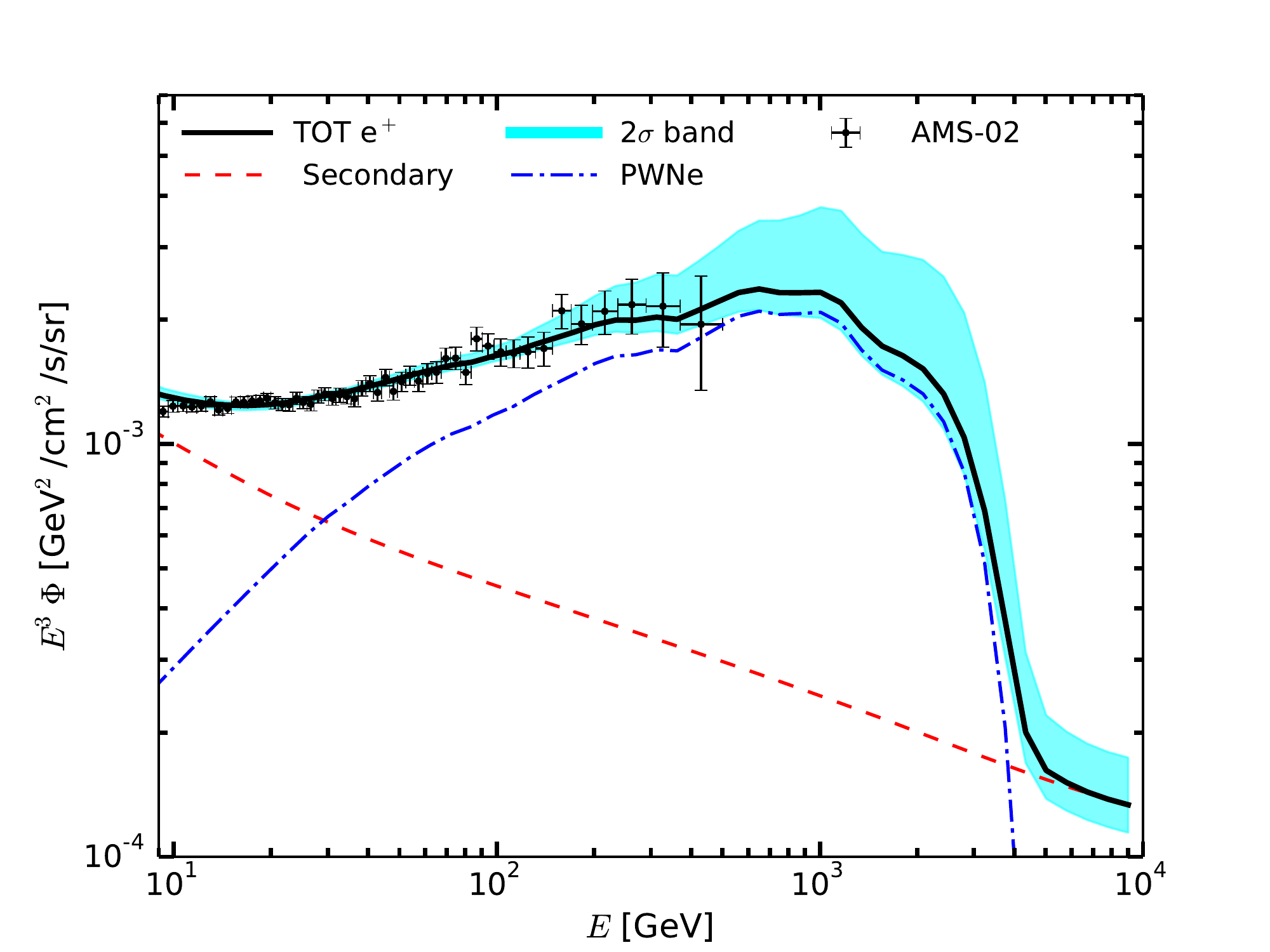}
\caption{Same as Fig.~\ref{fig:cut07kpc_case2}, but for the unknown SNR (black dotted line) discussed in {\it Case 3}. } 
\label{fig:cut07kpc_case3}
\end{figure*}
The results of the fit to AMS-02 data are shown in Fig.~\ref{fig:cut07kpc_case3} for both the $e^++e^-$ and $e^+$ data. The fit to the data 
is good on the whole energy range. Given the absence of other SNRs in the 0.7 kpc  around the Earth, included the high energy emitter Cygnus Loop, the wiggles by the PWNe are a bit more pronounced here than in Fig.~\ref{fig:cut07kpc_case2}. 
The unknown SNR turns out to be $t_s=144 \pm 9$ kyr old, be located at $d=0.36\pm0.06$ kpc from the Earth, to have a source index $\gamma=2.00\pm0.06$, and to shine with a total energy 
$E_{\rm tot} = 7.3\cdot10^{48}$ erg.  
Since this source is very old, the flux of electrons drops  at much lower energies than for Vela. The most energetic electrons have been slowed down by radiative cooling and diffusion in the long time since their injection. The role of the PWNe is such to keep explaining with high accuracy the AMS-02 positron spectrum.

\begin{figure*}
\includegraphics[width=0.52\textwidth]{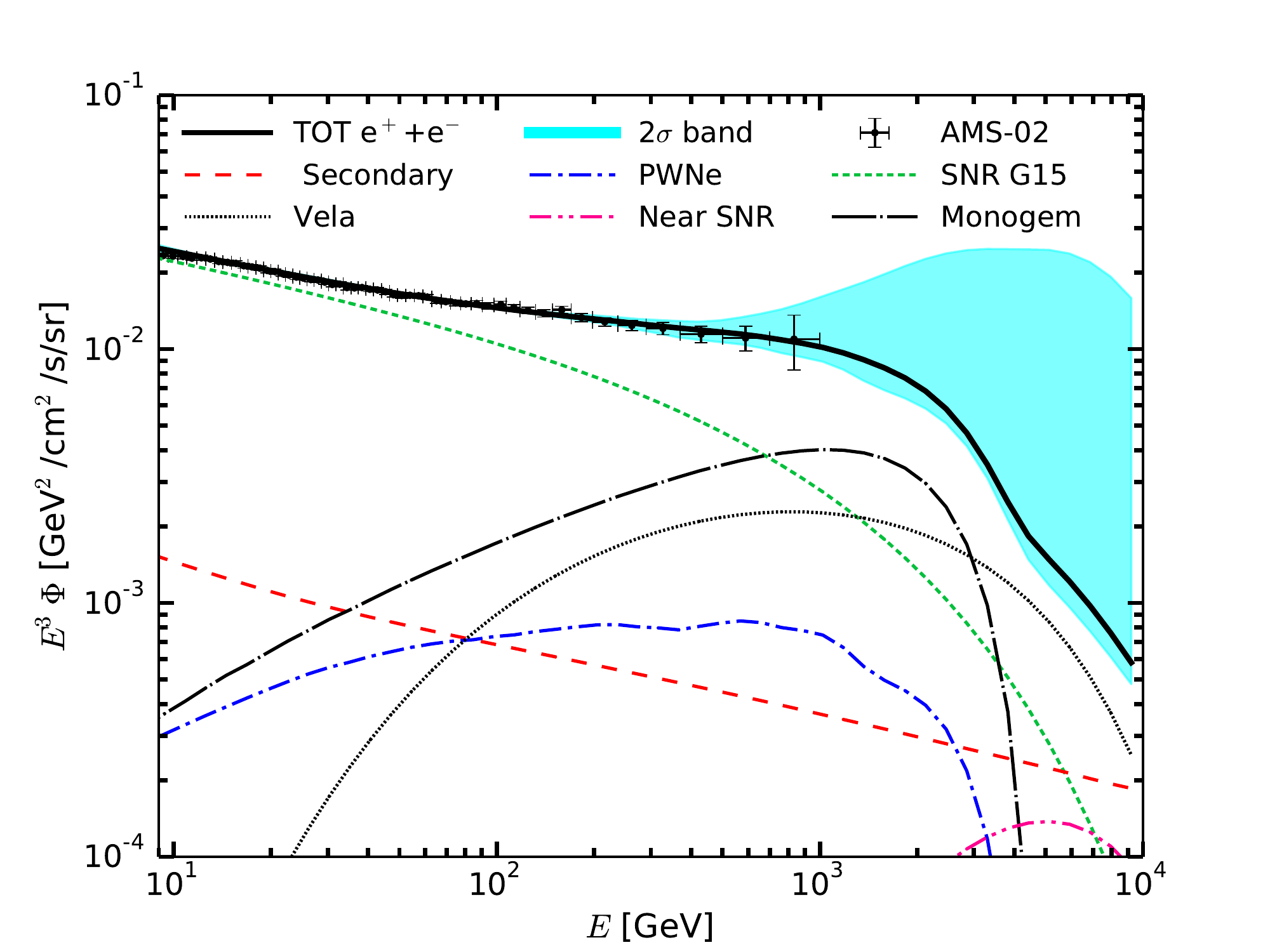}
\includegraphics[width=0.52\textwidth]{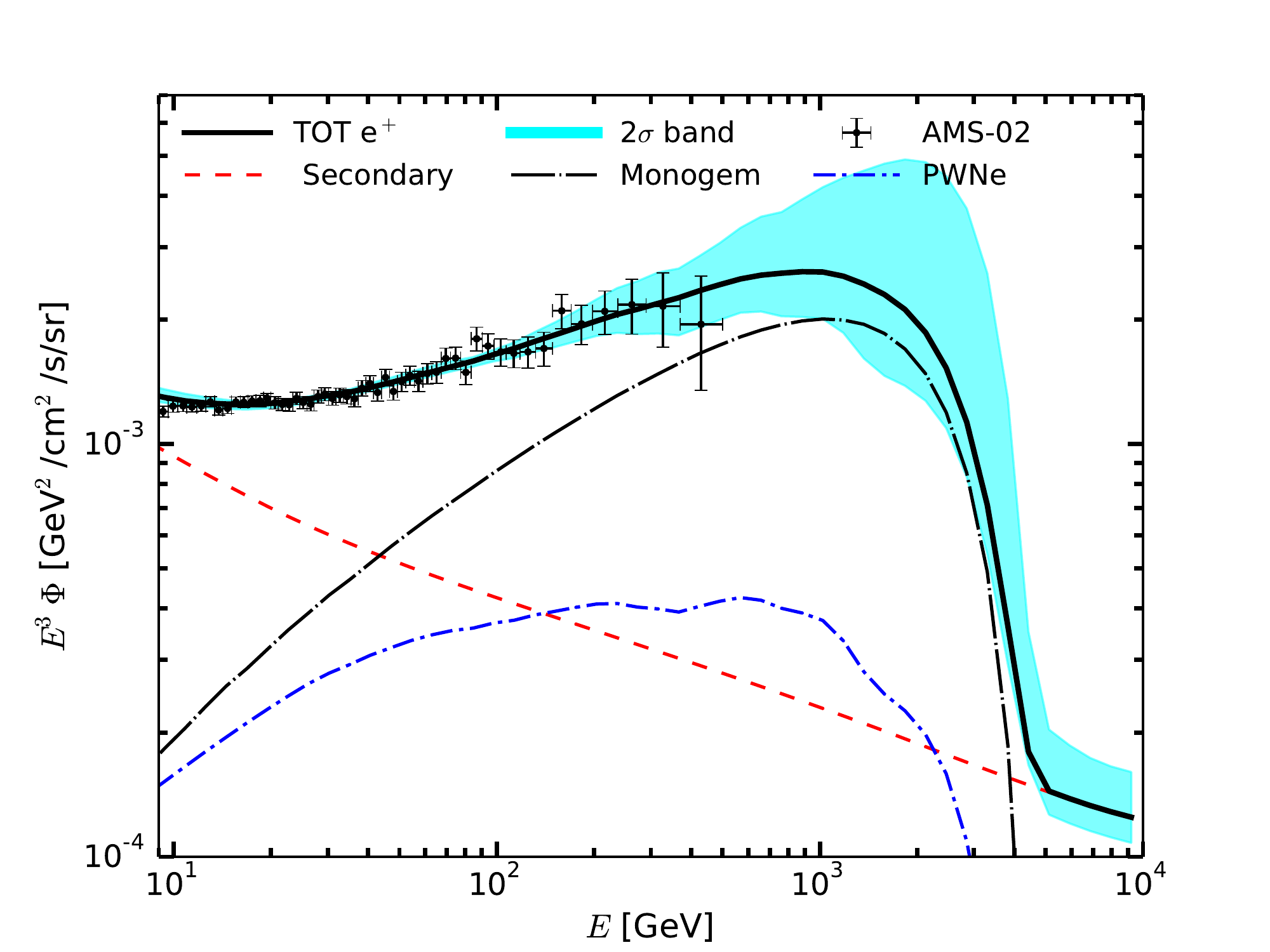}
\caption{Same as Fig.~\ref{fig:cut07kpc_case2}, but for dominant Monogem PWN (black double dot-dashed line) as discussed in {\it Case 4}. } 
\label{fig:flux_monogem}
\end{figure*}

\medskip
{\textit{\textbf{Case 4.} - Insights on close PWNe.}}
\\
Similarly to {\it Case 2}, we inspect also the characteristic parameters of Monogem and Geminga, which lie within 0.7 kpc around the Earth. 
They are included among the sources with the higher electron integrated fluxes in Tab.~\ref{tab:integratedflux}, and could explain easily almost alone the rising of positron fraction (see \cite{2014JCAP...04..006D}).
To this aim, we work with $R_{\rm cut}$ = 0.7 kpc and fit the AMS-02 data with the secondary component, the {\it far} and smooth SNRs and all the Green catalog SNRs 
within 0.7 kpc. As for the $e^\pm$ pairs from PWNe, we include Monogem {\it or} Geminga with free parameters as well as all the other  PWNe in the ATNF catalog. 
The age and distance of Monogem and Geminga are fixed to their catalog value, while we vary their spectral index and efficiency. 
The free parameters for this final analysis are thus  $\phi_F$, $\tilde{q}_{sec}$, $\gamma_{\rm PWN}$, $\eta$, $\gamma_{\rm Mon}$  (or $\gamma_{\rm Gem}$) and $\eta_{\rm Mon}$ (or $\eta_{\rm Gem}$), $N_{\rm near}$, $B_{\rm Vela}$,  $\gamma_{\rm Vela}$, $\gamma_{\rm SNR}$, $E_{\rm tot,SNR}$.
This analysis lets one close and bright PWN to dominate in the positron (and more mildly in the sum) flux, and to have a prominent role, if any, in the 
dipole anisotropy. 
The results for the fit to AMS-02  $e^++e^-$ and $e^+$ flux data are presented in Fig.~\ref{fig:flux_monogem} for the case of dominant Monogem.
The analysis with dominant Geminga leads to a very similar conclusion for the $e^++e^-$ flux. The positron flux shows a very good fit as well, with the difference that the flux from Geminga peaks at about 500 GeV, given its older age. 
We can see that indeed Monogem is the dominant source of $e^+$ and $e^-$ at the higher energies. In particular, the role of Monogem is 
prominent in explaining the positron flux at energies above 80-100 GeV. 
The spectral index for the dominant source is found to be $\gamma_{\rm Mon, Gem} \sim 1.9$, 
while the parameters for the SNR population, Vela and secondary component are similar to Tab.~\ref{tab:case2velabf}.
The most significant result on the free parameters is that the efficiency for Monogem  {\it or} Geminga points toward $\eta_{\rm Mon, Gem}\sim 1$,
while the $\eta$ for all the other PWNe in the catalog takes lower values of $\eta \sim 0.04-0.001$ with respect to the {\it Case 1,2,3}.
This means that the predicted efficiency for the release of the $e^\pm$ pairs from the PWN to the ISM has to be close to 100\%, 
in some tension to what it is predicted by theoretical models \citep{2011ASSP...21..624B, 2009ApJ...703.2051G}. 
However, uncertainties in the total spindown energy or in the pulsar spindown timescale $\tau_0$ (see Eqs.\ref{eq:EtotPWN}, \ref{eq:W0PWN}) 
can alter effectively the resulting efficiency up to a factor of 5-10 (see \cite{Mlyshev:2009twa}).


. 
\medskip
\section{Results on the anisotropy in \lowercase{$e^-$} and \lowercase{$e^+$}}
\label{sec:anisotropy}
A search for anisotropies in CR leptons was performed by PAMELA, AMS-02 and {\it Fermi}-LAT experiments, 
ending up in dipole anisotropy upper limits. 
The \textit{Fermi}-LAT experiment searched for an $e^++e^-$ anisotropy in the first year of data \citep{2010PhRvD..82i2003A}, for 
more than $1.6 \times 10^6$ particles with energies above $ 60$~GeV. This  threshold energy was chosen 
to minimize the influence of the geomagnetic field and of the Heliospheric Magnetic Field, both affecting the direction of detected charged particles in the GeV range. 
Upper limits on the $e^++e^-$ dipole anisotropy $\Delta_{e^+ +e^-}$ were obtained from a more general  analysis based on the spherical harmonics 
development. 
The data are provided in bins of energy integrated from a minimum energy $E_{\rm min}$, with $E_{\rm min}$  from 
$60$~GeV to $480$~GeV. 
The analysis includes a wide interval for the integration radii (from $\sim10^{\circ}$ to $90^{\circ}$).
The resulting upper limit on the dipole anisotropy increases from  $\Delta_{e^+ +e^-} \lesssim 0.005$ to $\Delta_{e^+ +e^-} \lesssim 0.10$ with increasing minimum energy. 
\\ 
An upper limit  on the positron to electron ratio dipole anisotropy $\Delta_{e^+/e^-}$ has been 
reported  by  AMS-02  for energies above $16$~GeV. The results are $\Delta_{e^+/e^-}\leq 0.036$ for the data in \cite{PhysRevLett.110.141102},  and $\Delta_{e^+/e^-}\leq 0.030$ for the higher statistics data set in  \cite{PhysRevLett.113.121101} at $95$\% C.L.. 
\\
The PAMELA Collaboration has performed a search on large-scale $e^+$ dipole anisotropy with the first four years of data \citep{Adriani:2015kfa}. 
The sample consists of $1489$ $e^+$ with rigidity $10\leq R\leq200$~GV. 
In order to account for the instrument exposure and other detector effects, the results are given in terms of positron over proton ratio
(the proton flux being considered isotropic, \cite{2013NuPhS.239..123G}). 
Compatibility with an isotropic distribution of positron arrival directions has been found, setting an  upper limit of $\Delta_{e^+}\leq 0.166$ at $95$\% C.L.. 
\\
Few remarks follow. First,  existing upper limits on lepton anisotropy from different experiments concern the flux of different observables: $e^++e^-$ for 
\textit{Fermi}-LAT, $e^+/e^-$ for AMS-02 and $e^+$ for PAMELA. While in the {\it Fermi}-LAT analysis the detected flux directional properties are compared with two different techniques of no-anisotropy map creation, 
the PAMELA and AMS-02 analysis are based on the comparison with the proton flux in the former case, and the electron flux in the latter case, both supposed to be isotropic.
Second, existing upper limits deal with \textit{integrate} dipole anisotropy as a function of a \textit{minimum} energy. 
Therefore, to properly compare theoretical predictions with experimental upper limits, 
we  integrate our predictions  from the indicated minimum energy $E_{\rm min}$ to a maximum energy $E_{\rm max}=5$~TeV. 
The integration is performed separately for the $\psi_{e^+ + e^-}^{s}(E)$ and the $\psi_{e^+ + e^-}^{tot}(E)$ terms in Eq. \ref{eq:eleposdipole}, following 
\cite{2010PhRvD..82i2003A}. Furthermore, we have verified that the results are unchanged if the integration of the term at numerator includes the energy dependent pre factor $K(E)/\lambda^2(E, E_{s})$. 
We will also present some result for the non-integrated dipole anisotropy as a function of the energy.

\subsection{The anisotropy from the Vela SNR}
\label{sec:velasnr}
\begin{figure}[h]
\centering
\includegraphics[width=0.7\textwidth]{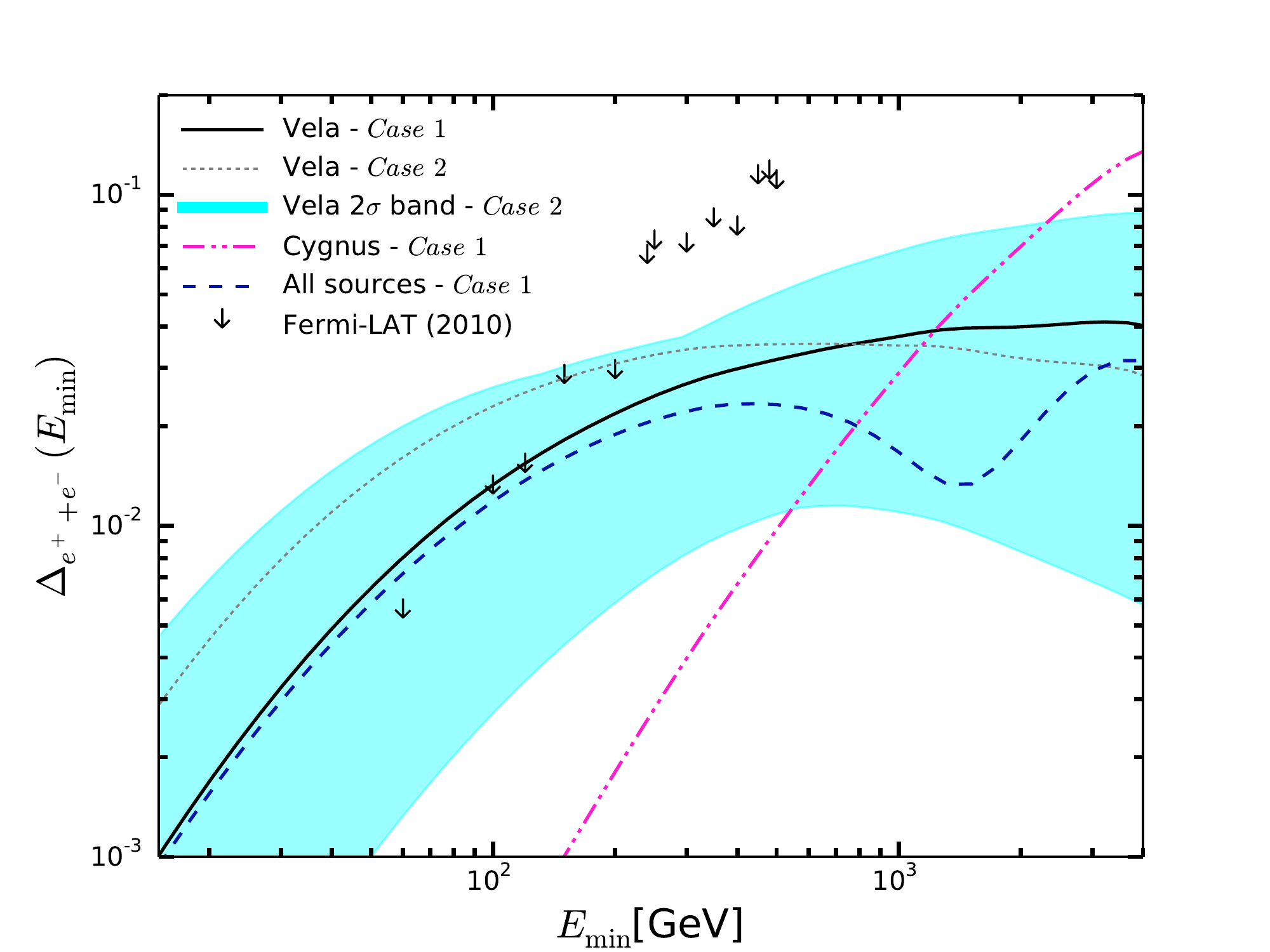}
\caption{Predictions for the dipole anisotropy  in the $e^++e^-$ flux from single SNRs, 
 for $R_{\rm cut}$=0.7 kpc and MAX propagation parameters. 
The energy bins are integrated in energy from $E_{\rm min}$ up to 5~TeV. 
The anisotropy from Vela and  from Cygnus Loop resulting from the best fit of \textit{Case 1} analysis (see text) are plotted as solid black and 
dot-dashed magenta lines, respectively. 
The results for the Vela best fit and  2-$\sigma$ uncertainty band within \textit{Case 2} are shown by the black dotted line and cyan band. 
The blue dashed curve corresponds to the anisotropy from all the single sources considered in \textit{Case 1}. 
 The downward arrows reproduce the  {\it Fermi}-LAT upper limits.} 
\label{fig:Vela_anisotropy}
\end{figure}

The dipole anisotropy for the Vela SNR has been computed for the \textit{Case 1} and  \textit{Case 2} configurations compatible with the AMS-02 data on the $e^++e^-$ and $e^+$ data. 
The results for the $e^++e^-$ anisotropy are summarized in Fig.~\ref{fig:Vela_anisotropy}  as a function of $E_{\rm min}$ energy, and compared to 
{\it Fermi}-LAT upper limits. We plot $\Delta_{e^+e^-}$ for the best fit on the AMS-02 data in \textit{Case 1}, for the Vela and Cygnus Loop SNRs. We also plot the 
 anisotropy for Vela  with its 2-$\sigma$ uncertainty band, for the analysis in \textit{Case 2} (see Fig.~\ref{fig:cut07kpc_case2}). 
The Vela anisotropy is an increasing function of $E_{\rm min}$, at least up to few hundreds GeV, and depending on the model parameters. 
Its is predicted  with an uncertainty of one order of magnitude or larger, and only mildly dependent on the bin of integrated energy.  
For $E_{\rm min}$ below 200 GeV, the {\it Fermi}-LAT upper limits lie in the Vela anisotropy band, while for higher energies the experimental 
data they are at least a factor of two higher than the maximal expected $\Delta_{e^+e^-}$.
We have checked that a lower energy cutoff for the Vela SNR of $E_c=2$~TeV would give a lower dipole anisotropy in the TeV range by about a factor 2. 
The {\it Fermi}-LAT upper limits have already the power to test  some of the models for Vela emitting $e^-$ compatible with AMS-02 $e^++e^-$ and $e^+$ data.
The experimental sensitivity to our models is maximal for the first (three, indeed) data points. The anisotropy for the best fit from \textit{Case 2} analysis is 
in some tension with  {\it Fermi}-LAT data for  $E_{\rm min} \lsim $ 120 GeV. 

We are dealing with fluxes above 10 GeV in order to get rid of the effects of the solar wind on the fluxes detected at the Earth. 
Heliospheric effects on the arrival distribution of local interstellar electrons and positrons, and thus on a possible dipole anisotropy, are to date little known. 
In particular, due to the uncertain structure of the heliospheric magnetic field, the results depend on the particular field modeling. 
Several works demonstrated with simulations of heliosphere that the propagation of CR in the heliosphere remains a diffusion-dominated process, and that drift imprints only second-order effects on the solar modulation of the spectrum \citep{Strauss20141015}. 

\begin{figure*}[t]
\centering
\includegraphics[width=0.8\textwidth]{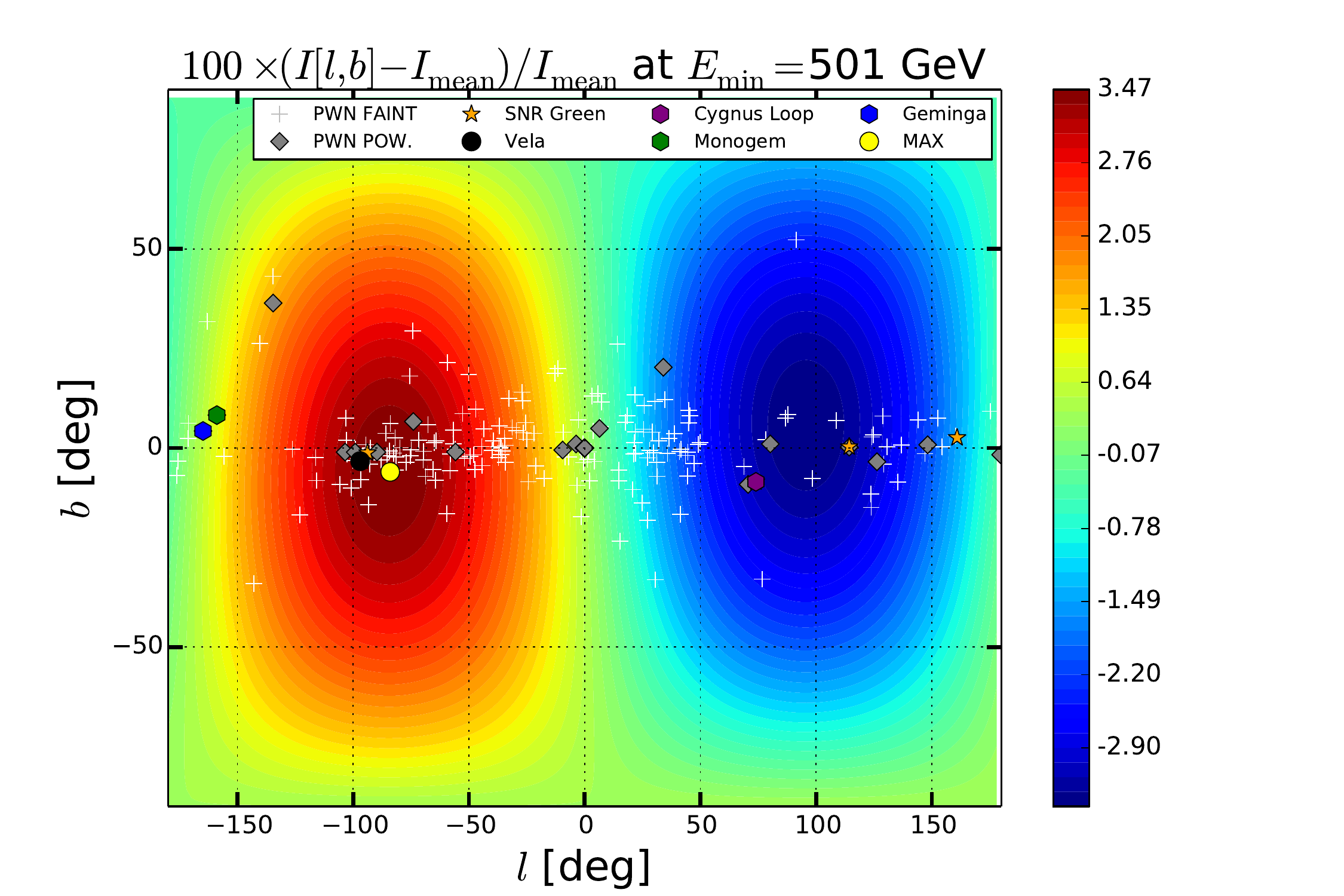} 
\includegraphics[width=0.8\textwidth]{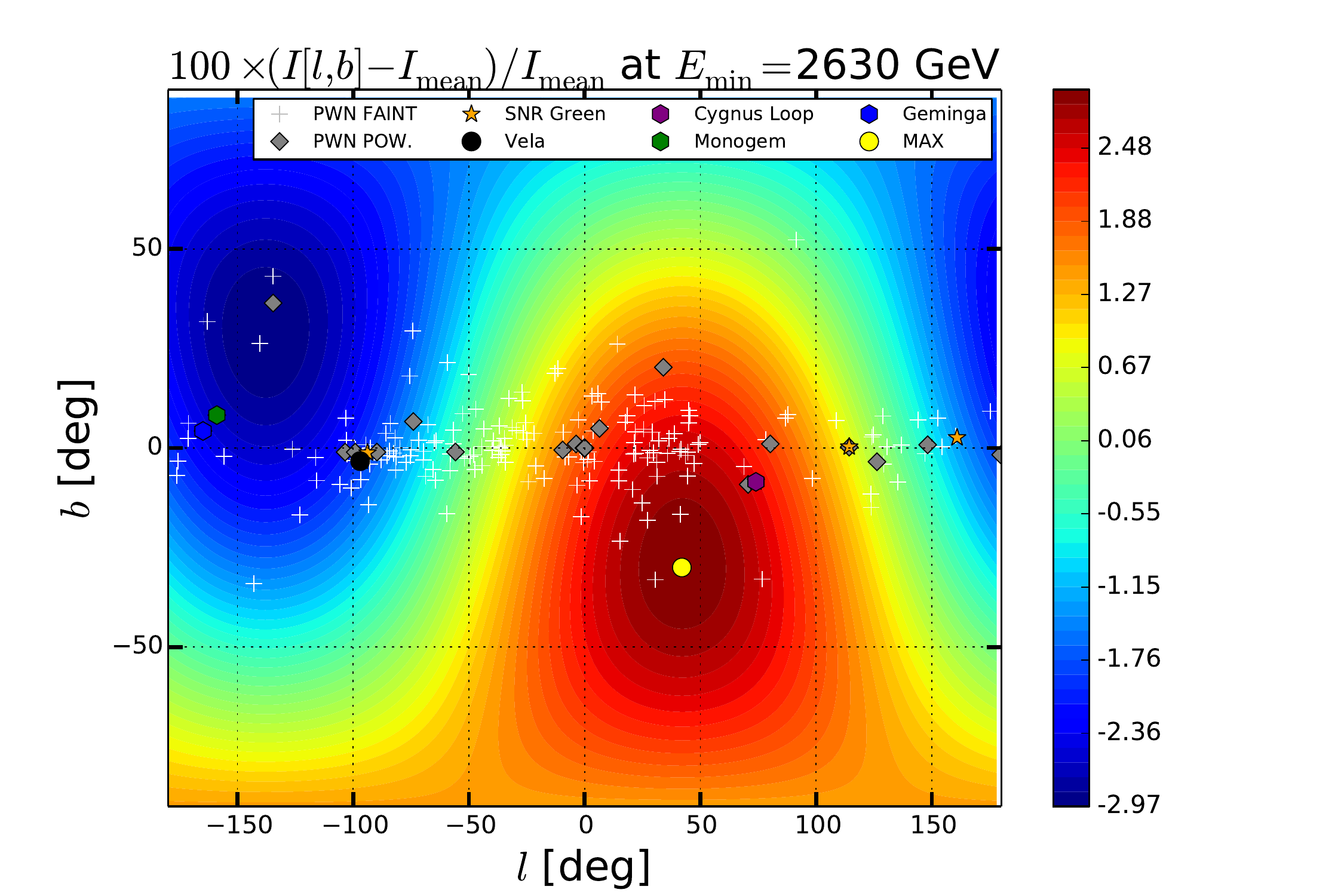} 
\caption{Contour plots of the intensity of the $e^++e^-$  flux for all the sources as treated in {\it Case 1}, as a function of the direction (l, b) [deg]. 
 We plot  the percentage relative difference between the intensity in any direction of sky and the mean intensity from the whole source collection, 
 for $E_{\rm min}$ = 501(2630) GeV in the top (bottom) panel. The yellow dot indicates the position of the maximum intensity, while the other symbols set the position of the sources, in particular  Vela (black dot), Cygnus (purple dot) and the SNRs in the Green catalog in the inner 0.7 kpc.  The color scale is linear and reflects the strength of the intensity in the given direction.}
\label{fig:contours}
 \end{figure*}

Future analysis of the whole  {\it Fermi}-LAT data set - now at its ninth year of operation -
could indeed hint at an anisotropy in the Vela direction, or set severe limits on a number of models selected by the inspection of flux data. 
Our analysis shows that the search for anisotropies in lepton data could be an interesting  complementary tool  when  local and powerful 
electron sources are invoked to shape the observed fluxes. 
\\
The composite anisotropy of all the single sources as included in {\it Case 1} analysis has been computed according to Eq.~\ref{eq:dipolesources}. 
The resulting anisotropy, in this specific model, shows a peculiar feature with two bumps, clearly tracking the dominant role of Vela and Cygnus SNRs in their energy domains. 
An insight on the role of Vela and Cygnus in the composition of the total anisotropy is given by the contour plots in Fig.~\ref{fig:contours}. 
We plot the percentage relative difference between the intensity in any direction of sky $I(l,b)$ ($l$ and $b$ being the Galactic longitude and the latitude) and the mean intensity from the whole source collection of sources. The interstellar $I(l,b)$ is computed according to Eq. 6 in \cite{1971ApL.....9..169S}, for the solutions to our diffusion equation 
(see Sect.~\ref{sec:thframe}). We show the results for  $E_{\rm min}$= 501 and $E_{\rm min}=$ 2630 GeV. The plots are a complementary visualization of the anisotropy from all the single sources shown in Fig.~\ref{fig:Vela_anisotropy} (blue dashed curve). 
At lower energy, the flux intensity is maximal in a direction very close to the Vela one. At higher energies, the maximal intensity shifts toward the Cygnus direction, with 
some offset driven by the collection of the other nearby sources. 

\subsection{The anisotropy from near PWNe} \label{anis_PWN}
\begin{figure*}[t]
\centering
\includegraphics[width=0.6\textwidth]{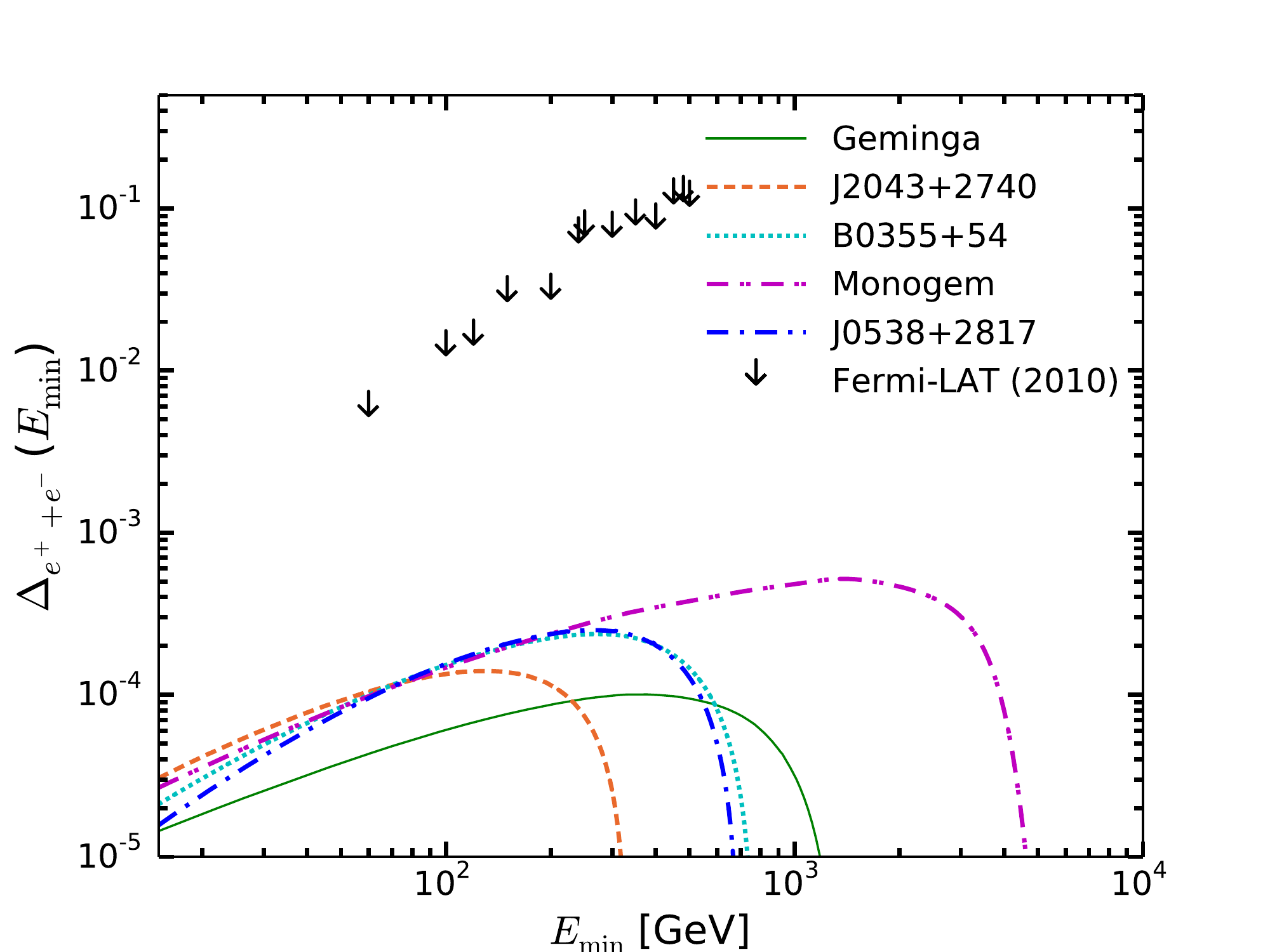}
\caption{Predictions for the anisotropy from single PWNe in the $e^++e^-$ flux   
along with experimental upper limits from {\it Fermi}-LAT, 
for $R_{\rm cut}$ = 0 kpc and MAX propagation parameters. 
The energy bins are integrated in energy. The results are for the five most powerful PWNe, as labeled inside the panels. }
\label{fig:PWN_anisotropy}
\end{figure*}

\begin{figure*}[t]
\includegraphics[width=0.52\textwidth]{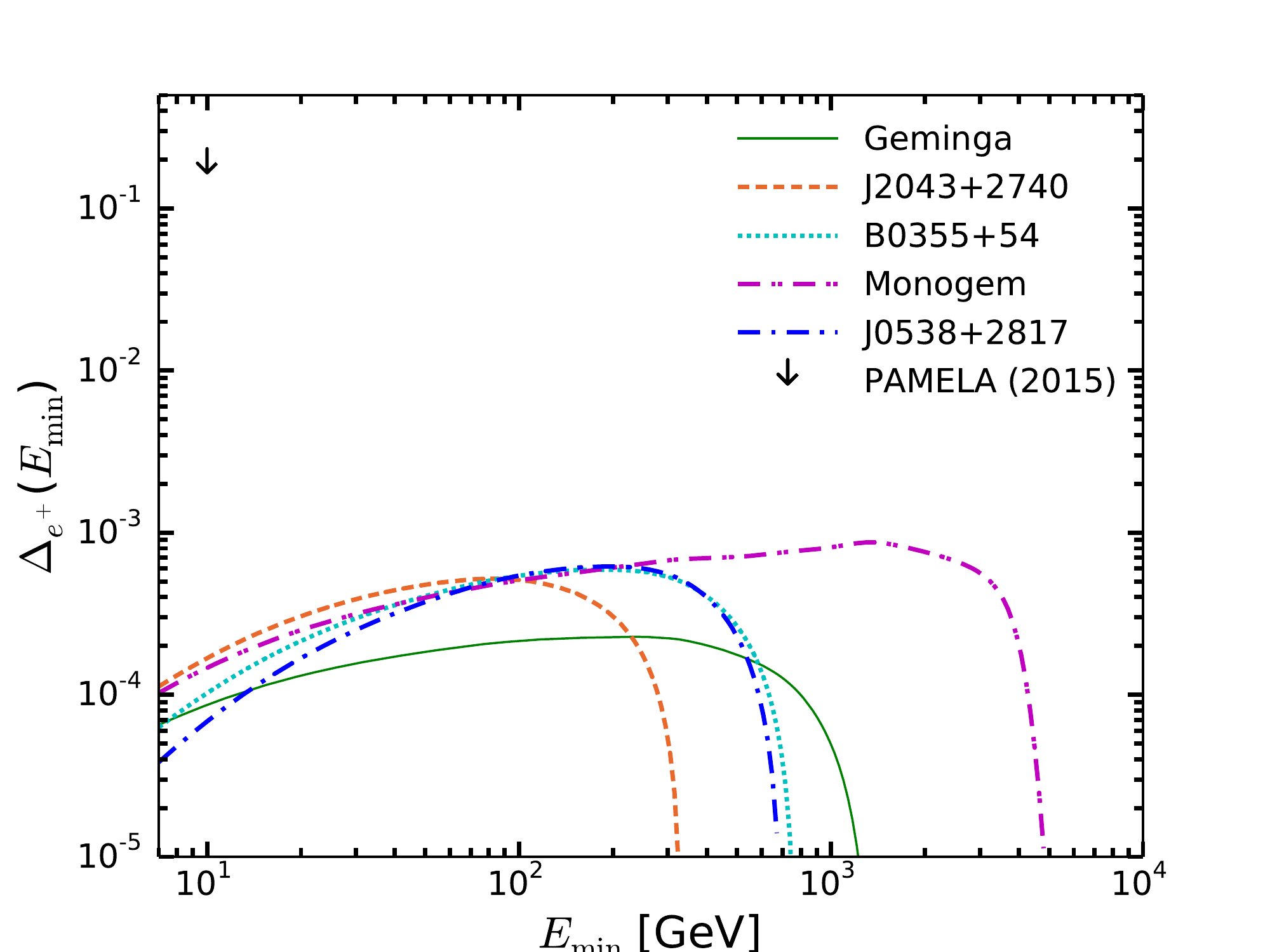} 
\includegraphics[width=0.52\textwidth]{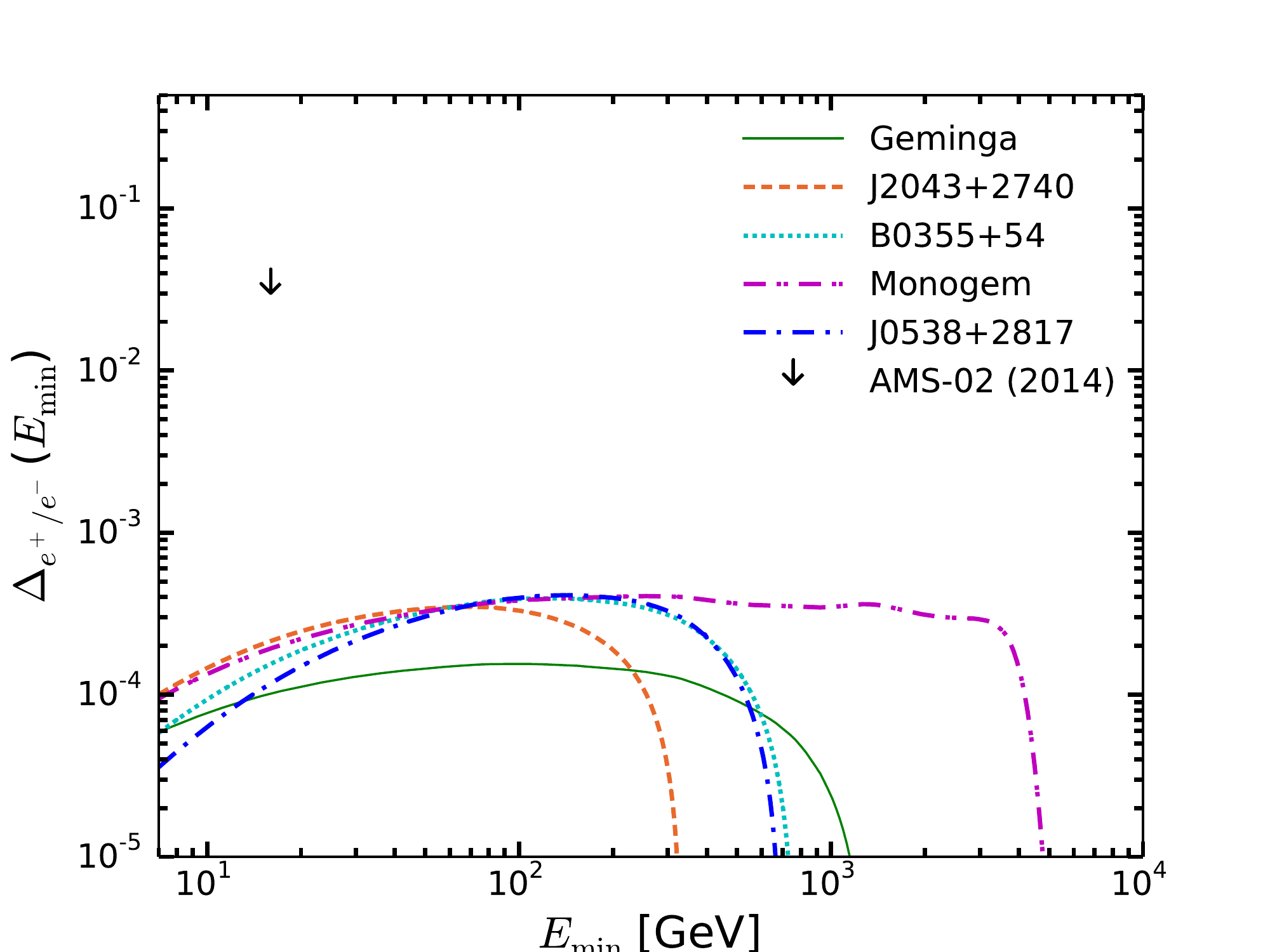}
\caption{Predictions for the anisotropy from single PWNe in the $e^+$ (left) and $e^+/e^-$ (right) fluxes, 
along with experimental upper limits from  Pamela (left) and AMS-02 (right), 
for $R_{\rm cut}$ = 0 kpc and MAX propagation parameters. 
The energy bins are integrated in energy. The results are for the five most powerful PWNe, as labeled inside the panels. }
\label{fig:PWN_anisotropy2}
\end{figure*}

The contribution to the $e^+$ and $e^-$ flux  from PWNe has been computed from sources taken directly from the ATNF catalog. We show here the results of the dipole anisotropy for the five most powerful PWNe in terms of the flux as identified in \cite{2014JCAP...04..006D}. 
The fluxes from each single PWN correspond to the best fits on AMS-02 data for the MAX scenario and $R_{\rm cut}$= 0~kpc, 
the SNRs contributing to $e^-$ as a smooth distribution of sources on the whole Galaxy.  
Our results are shown in Figs.~\ref{fig:PWN_anisotropy}, \ref{fig:PWN_anisotropy2}  for dipole anisotropy in $e^++e^-$, $e^+$ and $e^+/e^-$, along with the most recent relevant upper limits. 
None of the most powerful PWNe significantly dominates over the others, and all are predicted with a anisotropy level between 
$10^{-4}$ and  $10^{-3}$. The highest anisotropy is however reached by Monogem, for which a dipole at the level of a few times $10^{-4}$ is 
predicted well beyond 1 TeV.  The strongest upper limits are from the {\it Fermi}-LAT, so $\Delta_{e^++e^-}$ is at present the channel with the least gap between theory and data. The $\Delta_{e^+}$ Pamela upper limit of  $\Delta_{e^+}\leq 0.166$ for $E_{\rm min} > $ 10 GeV, stands almost 
three orders of magnitude above our predictions for Monogem. 
It is therefore evident that the properties of ATNF PWNe as emerging from AMS-02 flux data are very unlikely be tested by present or forthcoming 
data on the positron anisotropy. 
Our results differ from what is found for example in \citep{2009APh....32..140G,2011APh....34..528D}, where a much higher 
anisotropy in the $e^+$ and $e^-$ flux for the Monogem PWN was found. We observe that, differently from those works, 
 our analysis is constrained by the AMS-02 on the lepton flux and, which is most important, from the AMS-02 data for the positron fraction. The latter data 
at high energy bounds significantly the PWN component. We also notice that in the cited papers 
the efficiency for the Monogem PWN is $\eta \sim 35-40 \%$, while the contribution from the full collection of ATNF sources is subdominant with respect to the Monogem one. 
We inspect in more details the role of Monogem and Geminga with respect to the other PWNe in the {\it Case 4} analysis.

\subsection{The anisotropy from a close unknown SNR}
\label{sec:rqsnr}
\begin{figure}[!htb]
\centering
\includegraphics[width=0.6\textwidth]{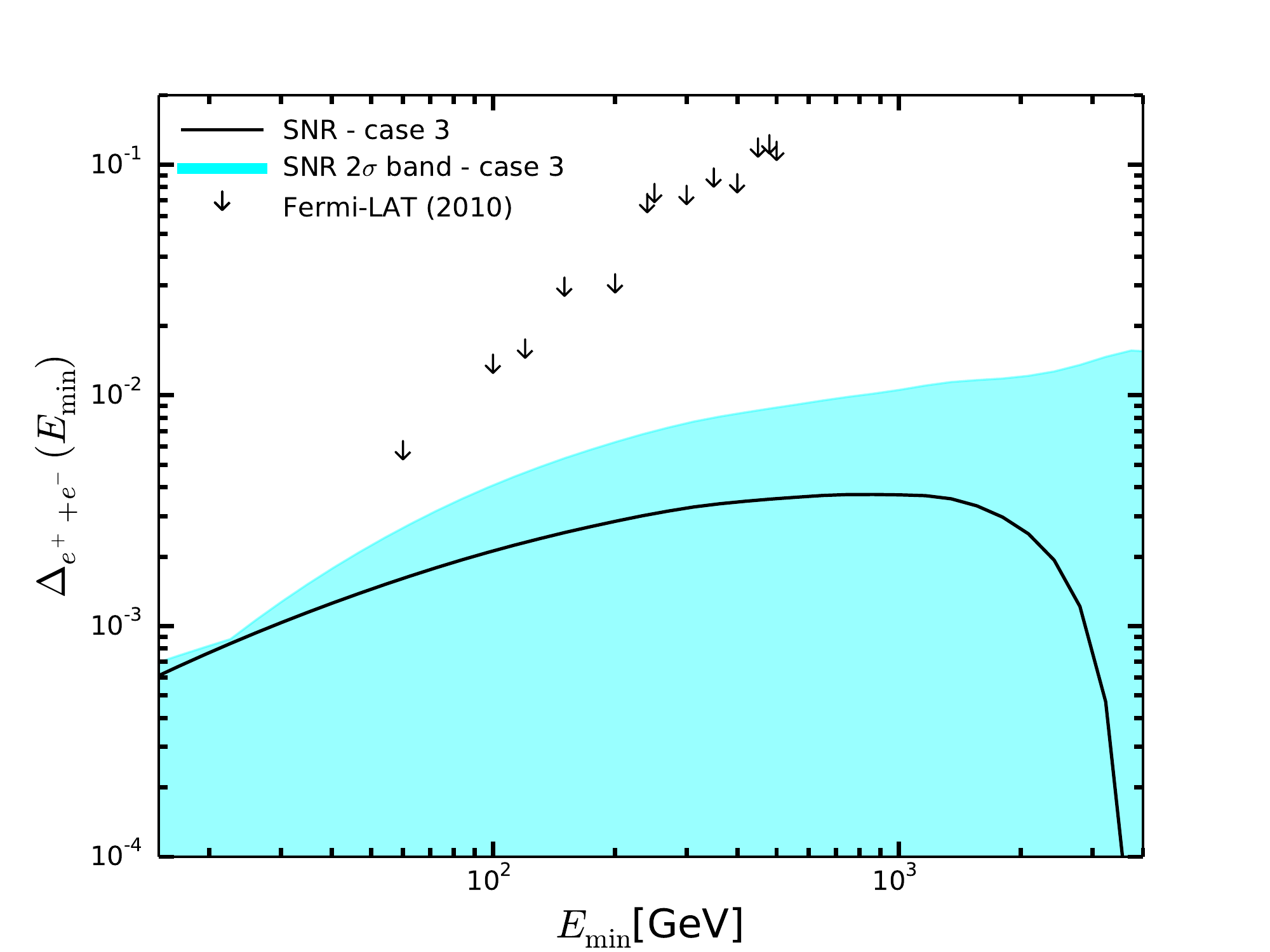}
\caption{Dipole anisotropy  in the $e^++e^-$ flux from the unknown SNRs
discussed in {\it Case 3}, for $R_{\rm cut}$=0.7 kpc and MAX propagation parameters. 
The results for the best fit and  2-$\sigma$ uncertainty band  are shown by the black dotted line and cyan band. 
 The downward arrows reproduce the  {\it Fermi}-LAT upper limits.} 
\label{fig:RQSNR_anisotropy}
\end{figure}
The anisotropy $\Delta_{e^++e^-}$ from a single, electromagnetically quiet SNR located at some radius $R< R_{\rm cut}$ = 0.7 kpc is plotted in Fig.~\ref{fig:RQSNR_anisotropy}. The line draws the result for the best fit and its 2-$\sigma$ uncertainty band to the AMS-02 flux data, under the {\it Case 3} hypothesis, as plotted in Fig.~\ref{fig:cut07kpc_case3}. 
The predicted anisotropy from this hypothetical SNR is lower than the Vela one (see Fig.~\ref{fig:Vela_anisotropy}), and a factor of two at least below current 
 {\it Fermi}-LAT data. Nevertheless, one could expect it to be explored by near future data. The 2-$\sigma$ uncertainty band is compatible with zero above $E_{\rm min}\sim150$~GeV. 
The behavior is driven by the remnant age, which for the best fit points to 144 kyr. 

\subsection{The anisotropy from Monogem and Geminga PWN}
\label{sec:ukpsr}
\begin{figure*}
\includegraphics[width=0.52\textwidth]{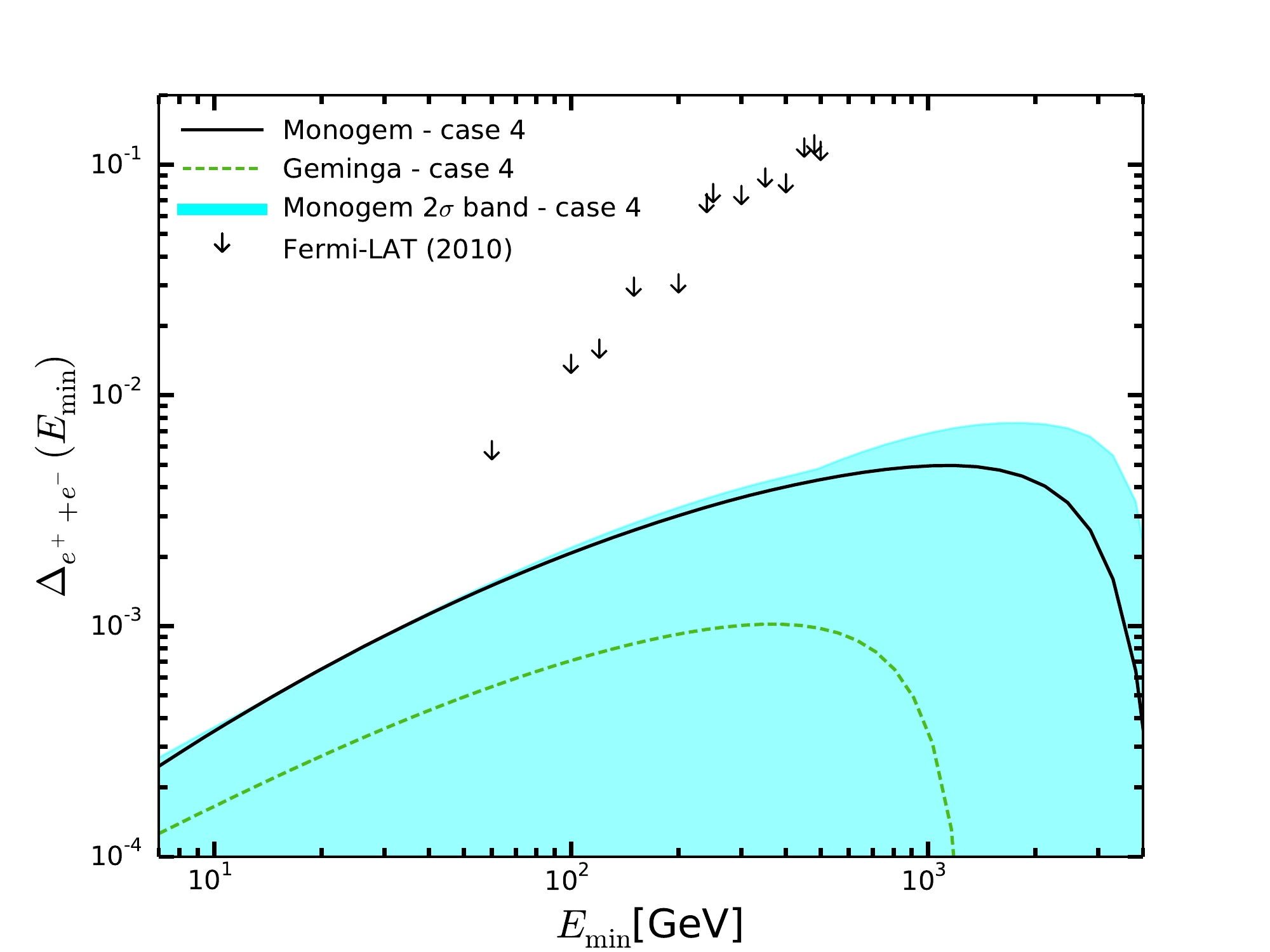}
\includegraphics[width=0.52\textwidth]{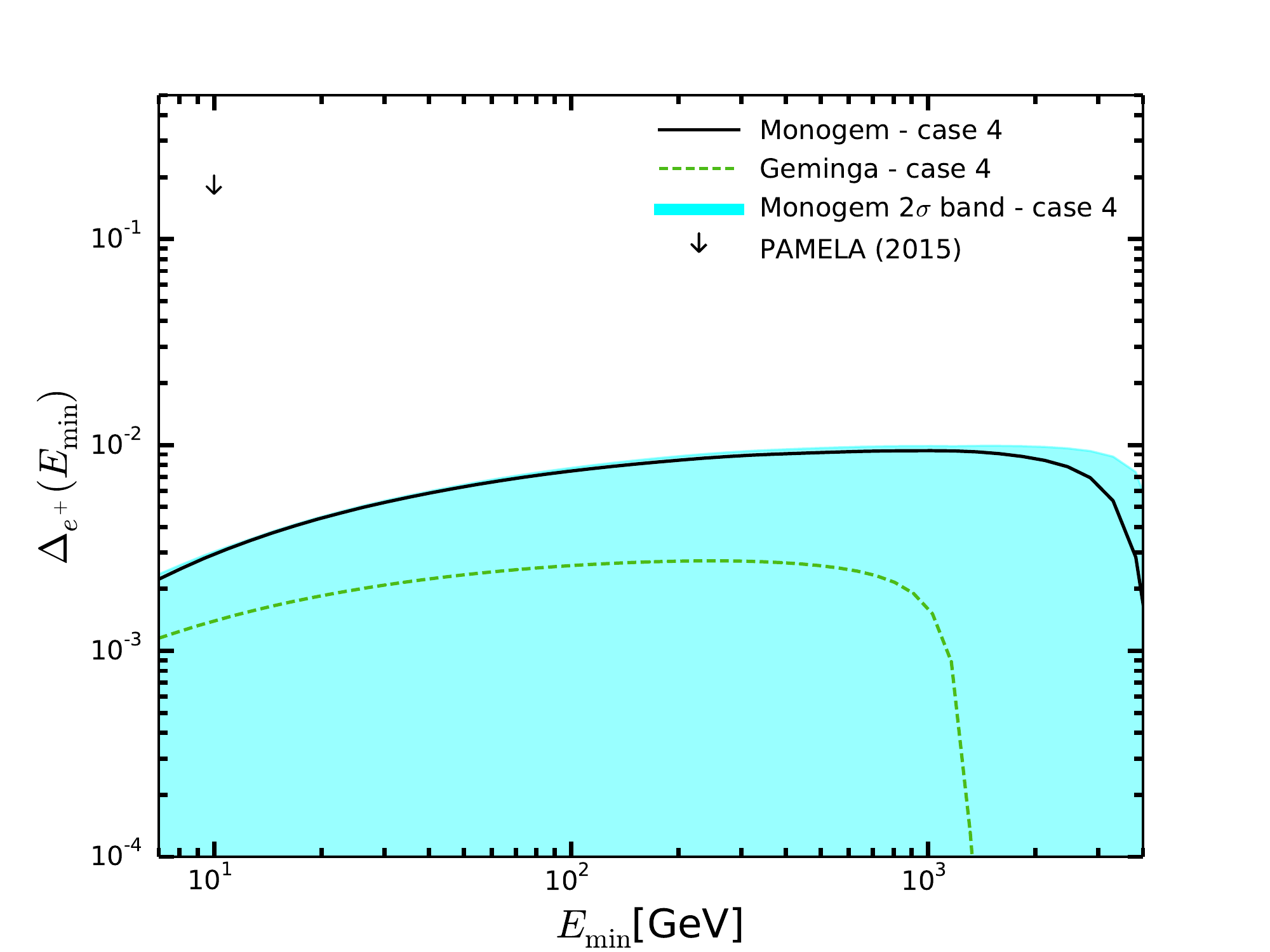}
\caption{Dipole anisotropy  in the $e^++e^-$ (left) and $e+$ (right)  flux from Monogem and Geminga as
discussed in {\it Case 4}, for $R_{\rm cut}$=0.7 kpc and MAX propagation parameters. 
The results for the best fit and  2-$\sigma$ uncertainty band  for Monogem are shown by the black line and cyan band, 
while the green dashed line indicates the best fit for Geminga. 
 The downward arrows reproduce the  {\it Fermi}-LAT upper limits.} 
\label{fig:mon_gem}
\end{figure*}

We report in Fig.~\ref{fig:mon_gem} the results for the dipole anisotropy from Monogem and Geminga, as derived from the {\it Case 4} fits. 
We plot the predictions for both the $\Delta_{e^++e^-}(E_{\rm min})$ and $\Delta_{e^+}(E_{\rm min})$. 
The comparison with Fig.~\ref{fig:PWN_anisotropy} shows that the Monogem anisotropy can be increased by up to a factor of ten, depending on the energies,  
with respect to the analysis in which it is treated democratically with all the other PWNe. This comment holds for both $e^++e^-$
 and the $e^+$ anisotropy. 
 Fig.~\ref{fig:mon_gem} also shows that the $\Delta_{e^++e^-}$ from Monogem  is about  a factor three (ten) below the current 
 {\it Fermi}-LAT upper limits at $E_{\rm min}=  60 (200)$ GeV. Also, the $\Delta_{e^+}$ is upshifted by about one order of magnitude with respect to the 
 study in Fig.~\ref{fig:PWN_anisotropy}, and stands two orders of magnitude below the current Pamela upper bounds.  Geminga gives results lower than a factor 2-3 with respect to Monogem, and decreasing rapidly to zero at lower energies of about 1 TeV, due to its age. 
 We have checked that a hypothetical (electromagnetically) unknown  PWN, consistent with the $e^+$ flux data, could  provide an anisotropy level greater than 
 a factor of a few than  the Monogem one, provided that it were even much closer to the Earth, and younger. 

\begin{figure*}
\includegraphics[width=0.52\textwidth]{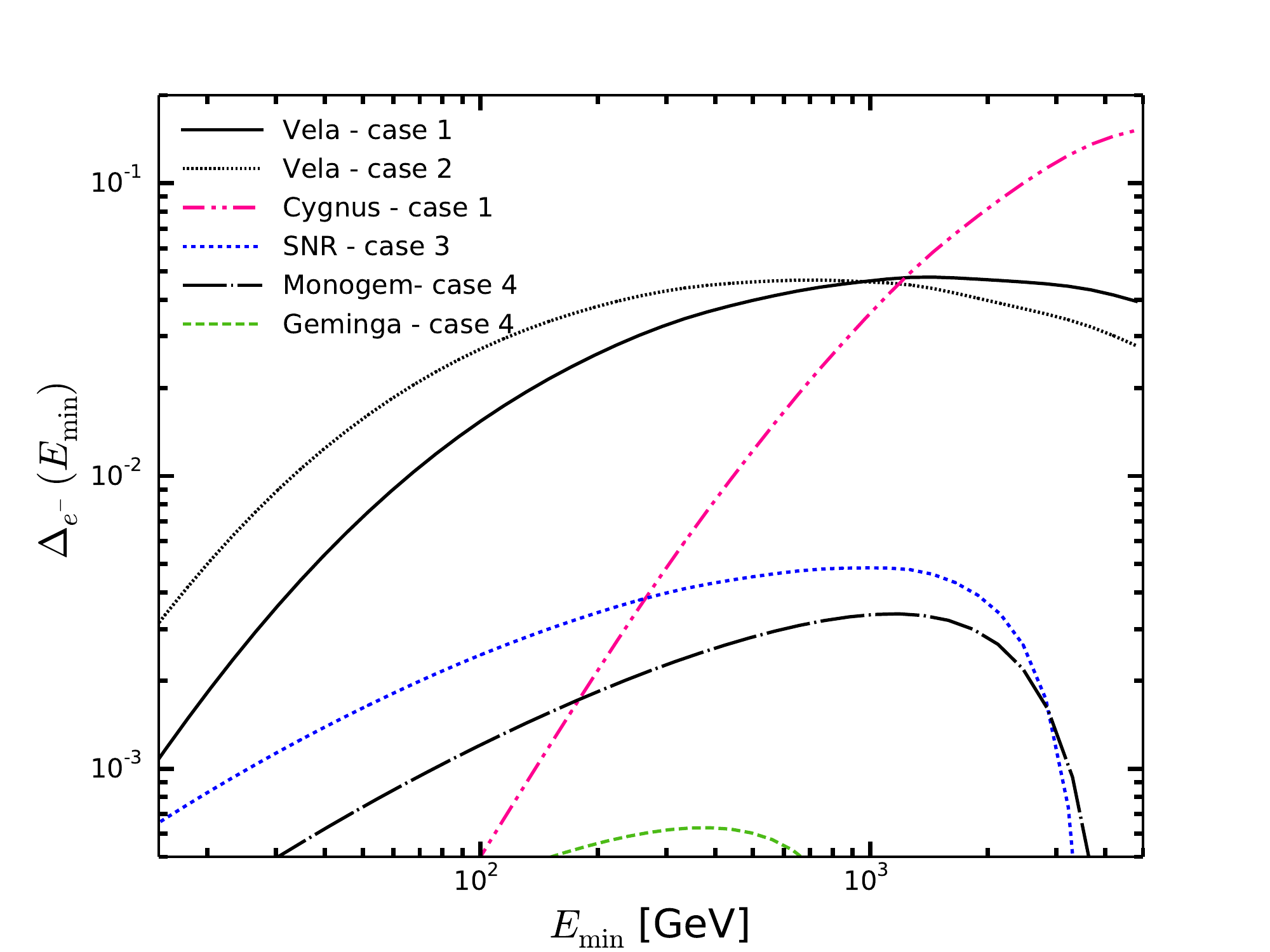}
\includegraphics[width=0.52\textwidth]{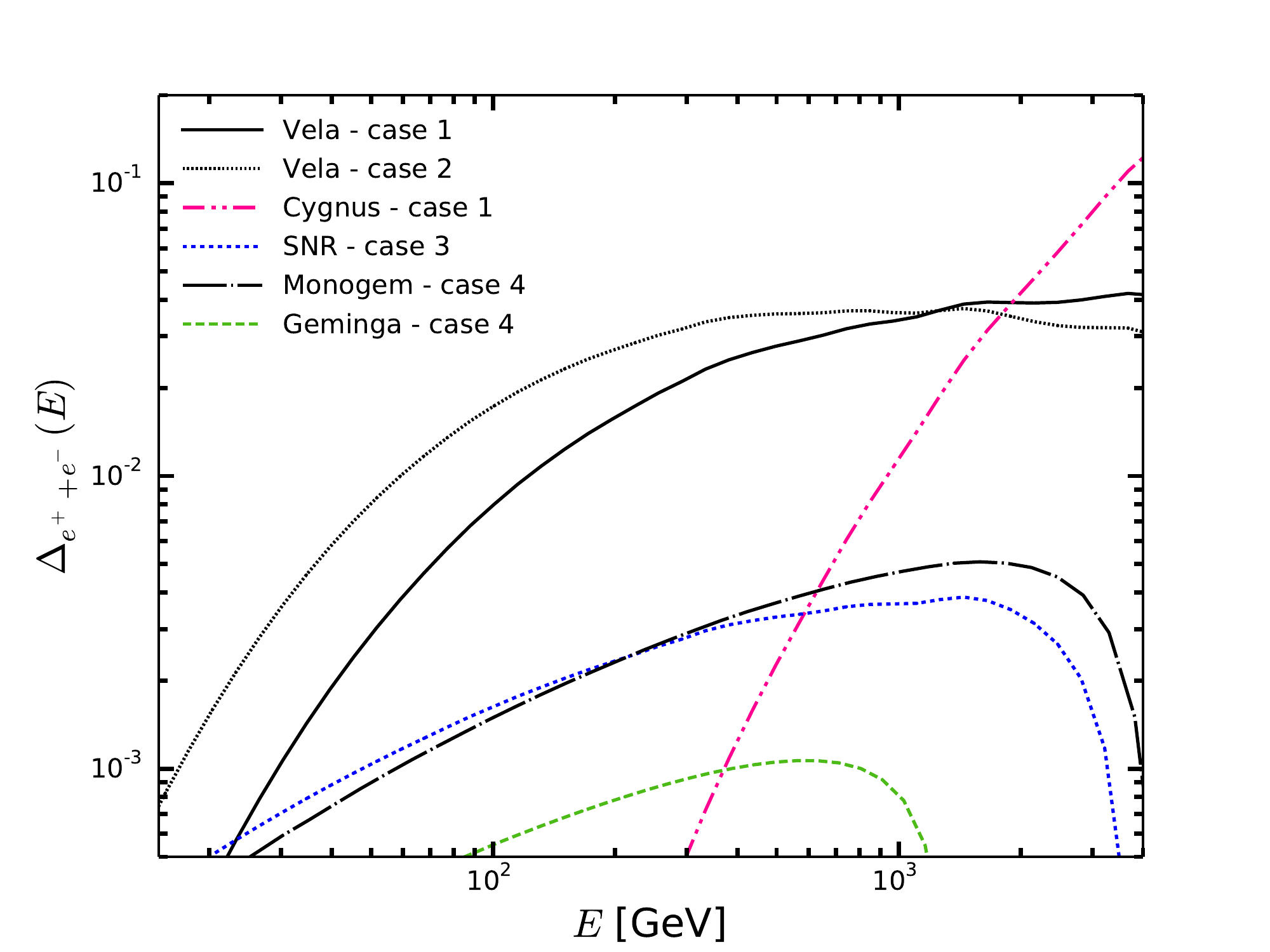}
\caption{Left panel: Dipole anisotropy in electrons  from Vela SNR as derived in {\it Case 1} and {\it Case 2}, Cygnus SNR as in {\it Case 1},  
the unknown SNR discussed in {\it Case 3}, and from the Monogem and the Geminga PWN  obtained in {\it Case 4} (see text for details), 
as a function of the minimum integrated energy $E_{\rm min}$.  
Right panel: Dipole anisotropy in electrons plus positrons for the same sources as in the left panel, but as a function of the energy $E$. 
} 
\label{fig:e_compilation}
\end{figure*}

In Fig.~\ref{fig:e_compilation}, we  plot the anisotropy in the electron flux  $\Delta_{e^-}(E_{\rm min})$ for the most significant single sources discussed up to now: Vela SNR as derived in {\it Case 1} and {\it Case 2}, Cygnus SNR as in {\it Case 1},  
the unknown SNR discussed in {\it Case 3}, and from the Monogem and the Geminga PWN  obtained in {\it Case 4}.
The case for $e^-$ sees Vela as a dominant source below integrated energies of about 1 TeV, as also found in the $e^++e^-$ observable. 
Monogem and Geminga  are predicted  below the unknown SNR level. 
We also investigate whether the search for an anisotropy as a function
of the energy $E$, and not integrated from a minimum energy $E_{\rm min}$, would provide better insights. The results are not very different, except for the role of Cygnus, which emerges at  few TeV, and the degeneracy between Monogem and the unknown SNR. Finally, due to a mere statistical effect, the low energy tail of the anisotropy spectrum is strongly depleted, with respect to the energy integral result. 

\medskip
\section{Conclusions}
\label{sec:concl}
We have discussed the phenomenology of a dipole anisotropy in the flux of electrons and positrons. 
We have tested several theoretical models  on the recent data from AMS-02 on the fluxes of $e^++e^-$ and $e^+$. 
All our predictions on the dipole anisotropy in the $e^++e^-$, $e^-$ and $e^+$ fluxes are consistent with the most recent AMS-02 data on the relevant fluxes. This is one major strength points of the present paper. 
In order to inspect more physically the role of local sources, we have shaped the Galaxy with a cut around the Earth, whose radius  $R_{\rm cut}$
has been tested at 3 and 0.7 kpc. As such, we have considered a smooth SNR population beyond  $R_{\rm cut}$, while the 
contribution inside $R_{\rm cut}$ comes from the single sources as directly found in the catalogs. 
Our model consists of: 
i)  an isotropic secondary $e^-$ and $e^+$ component, given by the fragmentation of  proton and helium CRs on the ISM;
ii) an isotropic $e^-$ flux injected by a smooth SNR population,
iii) an anisotropic production of $e^-$ from individual local SNRs, 
iv) an anisotropic injection of $e^\pm$ pairs from the PWNe in the ATNF catalog. 
The particles at sources have been propagated in a diffusion model, and properly treated for their strong radiative cooling. 
We have inspected different configurations for the contributions of local SNR and PWN to the electron and positron fluxes arriving at the Earth. 
In particular, we have studied the role of the Vela SNR, and of few PWNe. 

We have found that the anisotropy from the Vela SNR is at the level of the $e^++e^-$ {\it Fermi}-LAT upper limits obtained with one year data. The uncertainty on the dipole, as implied by AMS-02 flux data, is at least one order of magnitude wide. The composite anisotropy obtained from all the local sources is dominated by Vela below $E_{\rm min}$ of about 300 GeV, and by Cygnus at few TeV. 
We can expect that the next {\it Fermi}-LAT analysis in the $e^++e^-$ anisotropy data, performed in a much wider data set, starts to test some models 
for the Vela $e^-$ emission, now compatible with AMS-02 flux data. 
We have also explored the possibility that electrons are arriving at the Earth from a single, electromagnetically quiet SNR located at some radius $R< R_{\rm cut}$ = 0.7 kpc. The corresponding dipole anisotropy is a bit lower than the one obtained from Vela, mostly because of the much older 
age found for this source in the fit to AMS $e^++e^-$ data. 
We have also explored the dipole anisotropy from the most intense PWNe, with a dedicated analysis to Monogem and Geminga. 
The result is definitely lower than for the Vela SNR, even considering the dipole in the $e^+$ flux alone. 
The study for the anisotropy in the $e^-$ results in a dipole of order few $10^{-2}$ from the brightest SNRs: Vela at low energies, Cygnus above the TeV. 
We have finally checked the dipole as a function of the energy E, in addition to the analysis usually performed in terms of the minimum of the integrated energy. 
The results are not very different, except  the degeneracy between Monogem and the unknown SNR and the fact that, due to a mere statistical effect, the low energy tail of the anisotropy spectrum is strongly depleted.

As well as the forthcoming analysis on almost 9 years of data in the  {\it Fermi}-LAT data, 
ongoing experiments such as CALET and DAMPE will improve the flux energy resolution in the TeV region up to $\sim20$~TeV 
\citep{2011NIMPA.630...55T, 2015arXiv151208059M}. 
In particular, the CALET experiment includes in its main goals the observation of a possible anisotropy in the direction of nearby electron and positron sources, and could provide remarkable insights on the models tested in the present analysis. With data of this caliber, 
we can conclude our study claiming that the search of anisotropy in the lepton fluxes up to TeV energies 
can be an interesting tool for the inspection of properties of close SNRs, complementary to the high precision flux data.

\begin{acknowledgments}
We warmly thank Damiano Caprioli for interesting discussions and insights.
MDM acknowledges support by the NASA Fermi Guest Investigator Program 2014 through the Fermi 
multi-year Large Program N. 81303 (P.I. E.~Charles). 
\end{acknowledgments}

\bibliography{paper}

\providecommand{\href}[2]{#2}\begingroup\raggedright\begin{thebibliography}{10}

\bibitem{2009Natur.458..607A}
O.~Adriani et~al., {\it {An anomalous positron abundance in cosmic rays with
  energies 1.5-100GeV}},  {\em \nat} {\bf 458} (Apr., 2009) 607--609,
  [\href{http://xxx.lanl.gov/abs/0810.4995}{{\tt arXiv:0810.4995}}].

\bibitem{2012PhRvL.108a1103A}
M.~{Ackermann}, M.~{Ajello}, {Allafort}, et~al., {\it {Measurement of Separate
  Cosmic-Ray Electron and Positron Spectra with the Fermi Large Area
  Telescope}},  {\em Physical Review Letters} {\bf 108} (Jan., 2012) 011103,
  [\href{http://xxx.lanl.gov/abs/1109.0521}{{\tt arXiv:1109.0521}}].

\bibitem{PhysRevLett.110.141102}
{\bf AMS Collaboration} Collaboration, M.~Aguilar, G.~Alberti, B.~Alpat,
  Alvino, et~al., {\it First result from the alpha magnetic spectrometer on the
  international space station: Precision measurement of the positron fraction
  in primary cosmic rays of 0.5\char21{}350 gev},  {\em Phys. Rev. Lett.} {\bf
  110} (Apr, 2013) 141102.

\bibitem{PhysRevLett.113.121101}
{\bf (AMS Collaboration)} Collaboration, L.~Accardo, M.~Aguilar, D.~Aisa,
  B.~Alpat, A.~Alvino, G.~Ambrosi, K.~Andeen, L.~Arruda, N.~Attig,
  P.~Azzarello, A.~Bachlechner, et~al., {\it High statistics measurement of the
  positron fraction in primary cosmic rays of 0.5\char21{}500 gev with the
  alpha magnetic spectrometer on the international space station},  {\em Phys.
  Rev. Lett.} {\bf 113} (Sep, 2014) 121101.

\bibitem{2014PhRvL.113l1102A}
M.~{Aguilar}, D.~{Aisa}, A.~{Alvino}, G.~{Ambrosi}, K.~{Andeen}, L.~{Arruda},
  N.~{Attig}, P.~{Azzarello}, A.~{Bachlechner}, F.~{Barao}, et~al., {\it
  {Electron and Positron Fluxes in Primary Cosmic Rays Measured with the Alpha
  Magnetic Spectrometer on the International Space Station}},  {\em Physical
  Review Letters} {\bf 113} (Sept., 2014) 121102.

\bibitem{Adriani:2013uda}
{\bf PAMELA} Collaboration, O.~Adriani et~al., {\it {Cosmic-Ray Positron Energy
  Spectrum Measured by PAMELA}},  {\em Phys. Rev. Lett.} {\bf 111} (2013)
  081102, [\href{http://xxx.lanl.gov/abs/1308.0133}{{\tt arXiv:1308.0133}}].

\bibitem{2011PhRvL.106t1101A}
O.~{Adriani}, G.~C. {Barbarino}, {Bazilevskaya}, et~al., {\it {Cosmic-Ray
  Electron Flux Measured by the PAMELA Experiment between 1 and 625 GeV}},
  {\em Physical Review Letters} {\bf 106} (May, 2011) 201101,
  [\href{http://xxx.lanl.gov/abs/1103.2880}{{\tt arXiv:1103.2880}}].

\bibitem{2014PhRvL.113v1102A}
M.~{Aguilar}, D.~{Aisa}, B.~{Alpat}, A.~{Alvino}, G.~{Ambrosi}, K.~{Andeen},
  L.~{Arruda}, N.~{Attig}, P.~{Azzarello}, A.~{Bachlechner}, et~al., {\it
  {Precision Measurement of the (e$^{+}$+e$^{-}$) Flux in Primary Cosmic Rays
  from 0.5 GeV to 1 TeV with the Alpha Magnetic Spectrometer on the
  International Space Station}},  {\em Physical Review Letters} {\bf 113}
  (Nov., 2014) 221102.

\bibitem{Abdo:2009zk}
{\bf Fermi-LAT} Collaboration, A.~A. Abdo et~al., {\it {Measurement of the
  Cosmic Ray e+ plus e- spectrum from 20 GeV to 1 TeV with the Fermi Large Area
  Telescope}},  {\em Phys. Rev. Lett.} {\bf 102} (2009) 181101,
  [\href{http://xxx.lanl.gov/abs/0905.0025}{{\tt arXiv:0905.0025}}].

\bibitem{Boudaud:2014dta}
M.~Boudaud et~al., {\it {A new look at the cosmic ray positron fraction}},
  {\em Astron. Astrophys.} {\bf 575} (2015) A67,
  [\href{http://xxx.lanl.gov/abs/1410.3799}{{\tt arXiv:1410.3799}}].

\bibitem{Jin:2014ica}
H.-B. Jin, Y.-L. Wu, and Y.-F. Zhou, {\it {Cosmic ray propagation and dark
  matter in light of the latest AMS-02 data}},  {\em JCAP} {\bf 1509} (2015),
  no.~09 049, [\href{http://xxx.lanl.gov/abs/1410.0171}{{\tt
  arXiv:1410.0171}}].

\bibitem{Yuan:2013eja}
Q.~Yuan, X.-J. Bi, G.-M. Chen, Y.-Q. Guo, S.-J. Lin, and X.~Zhang, {\it
  {Implications of the AMS-02 positron fraction in cosmic rays}},  {\em
  Astropart. Phys.} {\bf 60} (2015) 1--12,
  [\href{http://xxx.lanl.gov/abs/1304.1482}{{\tt arXiv:1304.1482}}].

\bibitem{Lin:2014vja}
S.-J. Lin, Q.~Yuan, and X.-J. Bi, {\it {Quantitative study of the AMS-02
  electron/positron spectra: Implications for pulsars and dark matter
  properties}},  {\em Phys. Rev.} {\bf D91} (2015), no.~6 063508,
  [\href{http://xxx.lanl.gov/abs/1409.6248}{{\tt arXiv:1409.6248}}].

\bibitem{Mertsch:2014poa}
P.~Mertsch and S.~Sarkar, {\it {AMS-02 data confront acceleration of cosmic ray
  secondaries in nearby sources}},  {\em Phys. Rev.} {\bf D90} (2014) 061301,
  [\href{http://xxx.lanl.gov/abs/1402.0855}{{\tt arXiv:1402.0855}}].

\bibitem{Gaggero:2013nfa}
D.~Gaggero, L.~Maccione, D.~Grasso, G.~Di~Bernardo, and C.~Evoli, {\it {PAMELA
  and AMS-02 $e^+$ and $e^-$ spectra are reproduced by three-dimensional
  cosmic-ray modeling}},  {\em Phys. Rev.} {\bf D89} (2014) 083007,
  [\href{http://xxx.lanl.gov/abs/1311.5575}{{\tt arXiv:1311.5575}}].

\bibitem{Mlyshev:2009twa}
D.~Malyshev, I.~Cholis, and J.~Gelfand, {\it {Pulsars versus Dark Matter
  Interpretation of ATIC/PAMELA}},  {\em Phys. Rev.} {\bf D80} (2009) 063005,
  [\href{http://xxx.lanl.gov/abs/0903.1310}{{\tt arXiv:0903.1310}}].

\bibitem{2014JCAP...04..006D}
M.~{Di Mauro}, F.~{Donato}, N.~{Fornengo}, R.~{Lineros}, and A.~{Vittino}, {\it
  {Interpretation of AMS-02 electrons and positrons data}},  {\em \jcap} {\bf
  4} (Apr., 2014) 6, [\href{http://xxx.lanl.gov/abs/1402.0321}{{\tt
  arXiv:1402.0321}}].

\bibitem{2016JCAP...05..031D}
M.~{Di Mauro}, F.~{Donato}, N.~{Fornengo}, and A.~{Vittino}, {\it {Dark matter
  vs. astrophysics in the interpretation of AMS-02 electron and positron
  data}},  {\em \jcap} {\bf 5} (May, 2016) 031,
  [\href{http://xxx.lanl.gov/abs/1507.0700}{{\tt arXiv:1507.0700}}].

\bibitem{1998ApJ...509..212S}
A.~W. {Strong} and I.~V. {Moskalenko}, {\it {Propagation of Cosmic-Ray Nucleons
  in the Galaxy}},  {\em \apj} {\bf 509} (Dec., 1998) 212--228,
  [\href{http://xxx.lanl.gov/abs/astro-ph/9807150}{{\tt astro-ph/9807150}}].

\bibitem{Delahaye:2008ua}
T.~Delahaye, F.~Donato, N.~Fornengo, J.~Lavalle, R.~Lineros, P.~Salati, and
  R.~Taillet, {\it {Galactic secondary positron flux at the Earth}},  {\em
  Astron. Astrophys.} {\bf 501} (2009) 821--833,
  [\href{http://xxx.lanl.gov/abs/0809.5268}{{\tt arXiv:0809.5268}}].

\bibitem{Evoli:2016xgn}
C.~Evoli, D.~Gaggero, A.~Vittino, G.~Di~Bernardo, M.~Di~Mauro, A.~Ligorini,
  P.~Ullio, and D.~Grasso, {\it {Cosmic-ray propagation with DRAGON2: I.
  numerical solver and astrophysical ingredients}},
  \href{http://xxx.lanl.gov/abs/1607.0788}{{\tt arXiv:1607.0788}}.

\bibitem{2009APh....32..140G}
D.~{Grasso}, S.~{Profumo}, A.~W. {Strong}, L.~{Baldini}, R.~{Bellazzini}, E.~D.
  {Bloom}, J.~{Bregeon}, G.~{Di Bernardo}, D.~{Gaggero}, N.~{Giglietto},
  T.~{Kamae}, L.~{Latronico}, F.~{Longo}, M.~N. {Mazziotta}, A.~A. {Moiseev},
  A.~{Morselli}, J.~F. {Ormes}, M.~{Pesce-Rollins}, M.~{Pohl}, M.~{Razzano},
  C.~{Sgro}, G.~{Spandre}, and T.~E. {Stephens}, {\it {On possible
  interpretations of the high energy electron-positron spectrum measured by the
  Fermi Large Area Telescope}},  {\em Astroparticle Physics} {\bf 32} (Sept.,
  2009) 140--151, [\href{http://xxx.lanl.gov/abs/0905.0636}{{\tt
  arXiv:0905.0636}}].

\bibitem{2011APh....34..528D}
G.~{di Bernardo}, C.~{Evoli}, D.~{Gaggero}, D.~{Grasso}, L.~{Maccione}, and
  M.~N. {Mazziotta}, {\it {Implications of the cosmic ray electron spectrum and
  anisotropy measured with Fermi-LAT}},  {\em Astroparticle Physics} {\bf 34}
  (Feb., 2011) 528--538, [\href{http://xxx.lanl.gov/abs/1010.0174}{{\tt
  arXiv:1010.0174}}].

\bibitem{Hooper:2008kg}
D.~Hooper, P.~Blasi, and P.~D. Serpico, {\it {Pulsars as the Sources of High
  Energy Cosmic Ray Positrons}},  {\em JCAP} {\bf 0901} (2009) 025,
  [\href{http://xxx.lanl.gov/abs/0810.1527}{{\tt arXiv:0810.1527}}].

\bibitem{Kobayashi:2003kp}
T.~Kobayashi, Y.~Komori, K.~Yoshida, and J.~Nishimura, {\it {The most likely
  sources of high energy cosmic-ray electrons in supernova remnants}},  {\em
  Astrophys. J.} {\bf 601} (2004) 340--351,
  [\href{http://xxx.lanl.gov/abs/astro-ph/0308470}{{\tt astro-ph/0308470}}].

\bibitem{2010PhRvD..82i2003A}
M.~{Ackermann}, M.~{Ajello}, W.~B. {Atwood}, L.~{Baldini}, J.~{Ballet},
  G.~{Barbiellini}, D.~{Bastieri}, et~al., {\it {Searches for cosmic-ray
  electron anisotropies with the Fermi Large Area Telescope}},  {\em \prd} {\bf
  82} (Nov., 2010) 092003, [\href{http://xxx.lanl.gov/abs/1008.5119}{{\tt
  arXiv:1008.5119}}].

\bibitem{Adriani:2015kfa}
O.~Adriani et~al., {\it {Search for Anisotropies in Cosmic-ray Positrons
  Detected by the Pamela Experiment}},  {\em Astrophys. J.} {\bf 811} (2015),
  no.~1 21, [\href{http://xxx.lanl.gov/abs/1509.0624}{{\tt arXiv:1509.0624}}].

\bibitem{1971ApL.....9..169S}
C.~S. {Shen} and C.~Y. {Mao}, {\it {Anisotropy of High Energy Cosmic-Ray
  Electrons in the Discrete Source Model}},  {\em \aplett} {\bf 9} (Aug., 1971)
  169.

\bibitem{Buesching:2008hr}
I.~Buesching, O.~C. de~Jager, M.~S. Potgieter, and C.~Venter, {\it {A Cosmic
  Ray Positron Anisotropy due to Two Middle-Aged, Nearby Pulsars?}},  {\em
  Astrophys. J.} {\bf 678} (2008) L39--L42,
  [\href{http://xxx.lanl.gov/abs/0804.0220}{{\tt arXiv:0804.0220}}].

\bibitem{Pato:2010im}
M.~Pato, M.~Lattanzi, and G.~Bertone, {\it {Discriminating the source of
  high-energy positrons with AMS-02}},  {\em JCAP} {\bf 1012} (2010) 020,
  [\href{http://xxx.lanl.gov/abs/1010.5236}{{\tt arXiv:1010.5236}}].

\bibitem{2010APh....34...59C}
I.~{Cernuda}, {\it {Cosmic-ray electron anisotropies as a tool to discriminate
  between exotic and astrophysical sources}},  {\em Astroparticle Physics} {\bf
  34} (Sept., 2010) 59--69, [\href{http://xxx.lanl.gov/abs/0905.1653}{{\tt
  arXiv:0905.1653}}].

\bibitem{2010A&A...524A..51D}
T.~{Delahaye}, J.~{Lavalle}, R.~{Lineros}, F.~{Donato}, and N.~{Fornengo}, {\it
  {Galactic electrons and positrons at the Earth: new estimate of the primary
  and secondary fluxes}},  {\em \aap} {\bf 524} (Dec., 2010) A51,
  [\href{http://xxx.lanl.gov/abs/1002.1910}{{\tt arXiv:1002.1910}}].

\bibitem{Maurin:2001sj}
D.~Maurin, F.~Donato, R.~Taillet, and P.~Salati, {\it {Cosmic rays below z=30
  in a diffusion model: new constraints on propagation parameters}},  {\em
  Astrophys. J.} {\bf 555} (2001) 585--596,
  [\href{http://xxx.lanl.gov/abs/astro-ph/0101231}{{\tt astro-ph/0101231}}].

\bibitem{Gillessen:2008qv}
S.~Gillessen, F.~Eisenhauer, S.~Trippe, T.~Alexander, R.~Genzel, F.~Martins,
  and T.~Ott, {\it {Monitoring stellar orbits around the Massive Black Hole in
  the Galactic Center}},  {\em Astrophys. J.} {\bf 692} (2009) 1075--1109,
  [\href{http://xxx.lanl.gov/abs/0810.4674}{{\tt arXiv:0810.4674}}].

\bibitem{2007A&A...463..993S}
X.~H. {Sun}, J.~L. {Han}, W.~{Reich}, P.~{Reich}, W.~B. {Shi},
  R.~{Wielebinski}, and E.~{F{\"u}rst}, {\it {A Sino-German {$\lambda$}6 cm
  polarization survey of the Galactic plane. I. Survey strategy and results for
  the first survey region}},  {\em \aap} {\bf 463} (Mar., 2007) 993--1007,
  [\href{http://xxx.lanl.gov/abs/astro-ph/0611622}{{\tt astro-ph/0611622}}].

\bibitem{Donato:2003xg}
F.~Donato, N.~Fornengo, D.~Maurin, and P.~Salati, {\it {Antiprotons in cosmic
  rays from neutralino annihilation}},  {\em Phys. Rev.} {\bf D69} (2004)
  063501, [\href{http://xxx.lanl.gov/abs/astro-ph/0306207}{{\tt
  astro-ph/0306207}}].

\bibitem{Genolini:2015cta}
Y.~Genolini, A.~Putze, P.~Salati, and P.~D. Serpico, {\it {Theoretical
  uncertainties in extracting cosmic-ray diffusion parameters: the
  boron-to-carbon ratio}},  {\em Astron. Astrophys.} {\bf 580} (2015) A9,
  [\href{http://xxx.lanl.gov/abs/1504.0313}{{\tt arXiv:1504.0313}}].

\bibitem{Kappl:2015bqa}
R.~Kappl, A.~Reinert, and M.~W. Winkler, {\it {AMS-02 Antiprotons Reloaded}},
  {\em JCAP} {\bf 1510} (2015), no.~10 034,
  [\href{http://xxx.lanl.gov/abs/1506.0414}{{\tt arXiv:1506.0414}}].

\bibitem{2015MNRAS.454.1517G}
D.~A. {Green}, {\it {Constraints on the distribution of supernova remnants with
  Galactocentric radius}},  {\em \mnras} {\bf 454} (Dec., 2015) 1517--1524,
  [\href{http://xxx.lanl.gov/abs/1508.0293}{{\tt arXiv:1508.0293}}].

\bibitem{2004IAUS..218..105L}
D.~R. {Lorimer}, {\it {The Galactic Population and Birth Rate of Radio
  Pulsars}},  in {\em Young Neutron Stars and Their Environments} (F.~{Camilo}
  and B.~M. {Gaensler}, eds.), vol.~218 of {\em IAU Symposium}, p.~105, 2004.
\newblock \href{http://xxx.lanl.gov/abs/astro-ph/0308501}{{\tt
  astro-ph/0308501}}.

\bibitem{1998ApJ...504..761C}
G.~L. {Case} and D.~{Bhattacharya}, {\it {A New {$\Sigma$}-D Relation and Its
  Application to the Galactic Supernova Remnant Distribution}},  {\em \apj}
  {\bf 504} (Sept., 1998) 761--772,
  [\href{http://xxx.lanl.gov/abs/astro-ph/9807162}{{\tt astro-ph/9807162}}].

\bibitem{Gaensler:2006ua}
B.~M. Gaensler and P.~O. Slane, {\it {The evolution and structure of pulsar
  wind nebulae}},  {\em Ann. Rev. Astron. Astrophys.} {\bf 44} (2006) 17--47,
  [\href{http://xxx.lanl.gov/abs/astro-ph/0601081}{{\tt astro-ph/0601081}}].

\bibitem{1996ApJ...459L..83C}
X.~{Chi}, K.~S. {Cheng}, and E.~C.~M. {Young}, {\it {Pulsar Wind Origin of
  Cosmic Ray Positrons}},  {\em \apjl} {\bf 459} (Mar., 1996) L83.

\bibitem{2011ASSP...21..624B}
P.~{Blasi} and E.~{Amato}, {\it {Positrons from pulsar winds}},  {\em
  Astrophysics and Space Science Proceedings} {\bf 21} (2011) 624,
  [\href{http://xxx.lanl.gov/abs/1007.4745}{{\tt arXiv:1007.4745}}].

\bibitem{2005AJ....129.1993M}
R.~N. {Manchester}, G.~B. {Hobbs}, A.~{Teoh}, and M.~{Hobbs}, {\it {The
  Australia Telescope National Facility Pulsar Catalogue}},  {\em \aj} {\bf
  129} (Apr., 2005) 1993--2006,
  [\href{http://xxx.lanl.gov/abs/astro-ph/0412641}{{\tt astro-ph/0412641}}].

\bibitem{Green:2014cea}
D.~Green, {\it {A catalogue of 294 Galactic supernova remnants}},  {\em
  Bull.Astron.Soc.India} {\bf 42} (2014) 47,
  [\href{http://xxx.lanl.gov/abs/1409.0637}{{\tt arXiv:1409.0637}}].

\bibitem{Baltz:1998xv}
E.~A. Baltz and J.~Edsjo, {\it {Positron propagation and fluxes from neutralino
  annihilation in the halo}},  {\em Phys. Rev.} {\bf D59} (1998) 023511,
  [\href{http://xxx.lanl.gov/abs/astro-ph/9808243}{{\tt astro-ph/9808243}}].

\bibitem{Lavalle:2006vb}
J.~Lavalle, J.~Pochon, P.~Salati, and R.~Taillet, {\it {Clumpiness of dark
  matter and positron annihilation signal: computing the odds of the galactic
  lottery}},  {\em Astron. Astrophys.} {\bf 462} (2007) 827--848,
  [\href{http://xxx.lanl.gov/abs/astro-ph/0603796}{{\tt astro-ph/0603796}}].

\bibitem{Aharonian:1995zz}
F.~A. Aharonian, A.~M. Atoyan, and H.~J. Volk, {\it {High energy electrons and
  positrons in cosmic rays as an indicator of the existence of a nearby cosmic
  tevatron}},  {\em Astron. Astrophys.} {\bf 294} (1995) L41--L44.

\bibitem{1964Ginzburg}
V.~L. {Ginzburg} and S.~I. {Syrovatskii}, {\it {The Origin of Cosmic Rays}}, .

\bibitem{2001A&A...372..636A}
H.~{Alvarez}, J.~{Aparici}, J.~{May}, and P.~{Reich}, {\it {The radio spectral
  index of the Vela supernova remnant}},  {\em \aap} {\bf 372} (June, 2001)
  636--643.

\bibitem{1970ApJ...159L..35R}
P.~E. {Reichley}, G.~S. {Downs}, and G.~A. {Morris}, {\it {Time-of-Arrival
  Observations of Eleven Pulsars}},  {\em \apjl} {\bf 159} (Jan., 1970).

\bibitem{Reynolds:2011nk}
S.~P. Reynolds, B.~M. Gaensler, and F.~Bocchino, {\it {Magnetic fields in
  supernova remnants and pulsar-wind nebulae}},  {\em Space Sci. Rev.} {\bf
  166} (2012) 231--261, [\href{http://xxx.lanl.gov/abs/1104.4047}{{\tt
  arXiv:1104.4047}}].

\bibitem{2006A&A...447..937S}
X.~H. {Sun}, W.~{Reich}, J.~L. {Han}, P.~{Reich}, and R.~{Wielebinski}, {\it
  {New {$\lambda$}6 cm observations of the Cygnus Loop}},  {\em \aap} {\bf 447}
  (Mar., 2006) 937--947, [\href{http://xxx.lanl.gov/abs/astro-ph/0510509}{{\tt
  astro-ph/0510509}}].

\bibitem{Dodson:2003ai}
R.~Dodson, D.~Legge, J.~E. Reynolds, and P.~M. McCulloch, {\it {The Vela
  pulsar's proper motion and parallax derived from VLBI observations}},  {\em
  Astrophys. J.} {\bf 596} (2003) 1137--1141,
  [\href{http://xxx.lanl.gov/abs/astro-ph/0302374}{{\tt astro-ph/0302374}}].

\bibitem{Cha:1999pn}
A.~N. Cha, K.~R. Sembach, and A.~C. Danks, {\it {The distance to the vela
  supernova remnant}},  {\em Astrophys. J.} {\bf 515} (1999) L25,
  [\href{http://xxx.lanl.gov/abs/astro-ph/9902230}{{\tt astro-ph/9902230}}].

\bibitem{2001ApJS..134..283S}
R.~K. {Smith} and D.~P. {Cox}, {\it {Multiple Supernova Remnant Models of the
  Local Bubble and the Soft X-Ray Background}},  {\em \apjs} {\bf 134} (June,
  2001) 283--309.

\bibitem{Helmut_LB_2011}
H.~A. Abt, {\it The age of the local interstellar bubble},  {\em The
  Astronomical Journal} {\bf 141} (2011), no.~5 165.

\bibitem{2009ApJ...703.2051G}
J.~D. {Gelfand}, P.~O. {Slane}, and W.~{Zhang}, {\it {A Dynamical Model for the
  Evolution of a Pulsar Wind Nebula Inside a Nonradiative Supernova Remnant}},
  {\em \apj} {\bf 703} (Oct., 2009) 2051--2067,
  [\href{http://xxx.lanl.gov/abs/0904.4053}{{\tt arXiv:0904.4053}}].

\bibitem{2013NuPhS.239..123G}
U.~{Giaccari}, O.~{Adriani}, {Barbarino}, et~al., {\it {Anisotropy studies in
  the cosmic ray proton flux with the PAMELA experiment}},  {\em Nuclear
  Physics B Proceedings Supplements} {\bf 239} (June, 2013) 123--128.

\bibitem{Strauss20141015}
R.~Strauss and M.~Potgieter, {\it Where does the heliospheric modulation of
  galactic cosmic rays start?},  {\em Advances in Space Research} {\bf 53}
  (2014), no.~7 1015 -- 1023.

\bibitem{2011NIMPA.630...55T}
S.~{Torii} and {CALET Collaboration}, {\it {Calorimetric electron telescope
  mission. Search for dark matter and nearby sources}},  {\em Nuclear
  Instruments and Methods in Physics Research A} {\bf 630} (Feb., 2011) 55--57.

\bibitem{2015arXiv151208059M}
P.~S. {Marrocchesi}, {\it {CALET: a high energy astroparticle physics
  experiment on the ISS}},  {\em ArXiv e-prints, 1512.08059} (Dec., 2015)
  [\href{http://xxx.lanl.gov/abs/1512.0805}{{\tt arXiv:1512.0805}}].

\end{thebibliography}\endgroup

\end{document}